\documentclass[aps,prc,reprint,superscriptaddress,showpacs,amsmath,amssymb,longbibliography,lengthcheck]{revtex4-1}

\usepackage[dvipdfmx]{graphicx}
\usepackage{mathrsfs}
\usepackage{txfonts}
\begin{document}

 \title{High-spin torus isomers and their precession motions}

\author{T. Ichikawa}%
\affiliation{Yukawa Institute for Theoretical Physics, Kyoto University,
Kyoto 606-8502, Japan}
\author{K. Matsuyanagi}
\affiliation{Yukawa Institute for Theoretical Physics, Kyoto University,
Kyoto 606-8502, Japan}
\affiliation{RIKEN Nishina Center, Wako 351-0198, Japan}
\author{J. A. Maruhn}
\affiliation{Institut fuer Theoretische Physik, 
Universitaet Frankfurt, D-60438 Frankfurt, Germany}
\author{N. Itagaki}
\affiliation{Yukawa Institute for Theoretical Physics, Kyoto University,
Kyoto 606-8502, Japan}
\date{\today}

\begin{abstract}
 \begin{description}
  \item[Background] In our previous study, we found that an exotic
	     isomer with a torus shape may exist in the high-spin, highly
	     excited states of $^{40}$Ca. The $z$ component of the total
	     angular momentum, $J_z=60$ $\hbar$, of this torus isomer is
	     constructed by aligning totally twelve single-particle
	     angular momenta in the direction of the symmetry axis of
	     the density distribution.  The torus isomer executes
	     precession motion with the rigid-body moments of inertia
	     about an axis perpendicular to the symmetry axis.  The
	     investigation, however, has been focused only on $^{40}$Ca.
  \item[Purpose] We systematically investigate the existence of exotic
	     torus isomers and their precession motions for a series of
	     $N=Z$ even-even nuclei from $^{28}$Si to $^{56}$Ni. We
	     analyze the microscopic shell structure of the torus isomer
	     and discuss why the torus shape is generated beyond the
	     limit of large oblate deformation.
  \item[Method] We use the cranked three-dimensional Hartree-Fock (HF)
	     method with various Skyrme interactions in a systematic
	     search for high-spin torus isomers. We use the
	     three-dimensional time-dependent Hartree-Fock (TDHF) method
	     for describing the precession motion of the torus isomer.
  \item[Results] We obtain high-spin torus isomers in $^{36}$Ar,
	     $^{40}$Ca, $^{44}$Ti, $^{48}$Cr, and $^{52}$Fe.  The
	     emergence of the torus isomers is associated with the
	     alignments of single-particle angular momenta, which is the
	     same mechanism as found in $^{40}$Ca.  It is found that all
	     the obtained torus isomers execute the precession motion at
	     least two rotational periods.  The moment of inertia about
	     a perpendicular axis, which characterizes the precession
	     motion, is found to be close to the classical rigid-body
	     value.
  \item[Conclusions] The high-spin torus isomer of $^{40}$Ca is not an
	     exceptional case. Similar torus isomers exist widely in
	     nuclei from $^{36}$Ar to $^{52}$Fe and they execute the
	     precession motion.  The torus shape is generated beyond the
	     limit of large oblate deformation by eliminating the $0s$
	     components from all the deformed single-particle wave
	     functions to maximize their mutual overlaps.
 \end{description}
 \end{abstract}


\maketitle

\section{Introduction}
Nuclear rotation is a key phenomenon to study the fundamental properties
of finite many-body quantum systems.  In particular, the rotation about
the symmetry ($z$) axis produces a unique quantum object with its
density distribution of a torus shape, as shown in our previous studies
for $^{40}$Ca \cite{ich12,ich14}.  In a classical picture for such
rotation the oblate deformation develops with increasing rotational
frequency due to the strong centrifugal force~\cite{Cohen}.  However,
such a collective rotation about the symmetry axis is
quantum-mechanically forbidden.  Instead, it is possible to construct
extremely high-spin states by aligning individual angular momenta of
single-particle motion in the direction of the symmetry
axis~\cite{BM77,Afan99}.

 A drastic example is a high-spin torus isomer in $^{40}$Ca
\cite{ich12}, where totally twelve single particles with the orbital
angular momenta $\Lambda=+4$, $+5$, and $+6$ align in the direction of
the symmetry axis and construct a $z$ component of the total angular
momentum of $J_z=60$ $\hbar$.  Thus, a ``macroscopic'' amount of
circulating current emerges in the torus isomer state, which may be
regarded as a fascinating new form of the nuclear matter suggested by
Bohr and Mottelson~\cite{BM81}.

Another important kind of rotation is a collective motion that restores
the symmetry spontaneously broken in the self-consistent mean field.
The density distribution of the torus isomer largely breaks the symmetry
about an ($x$ or $y$) axis perpendicular to the symmetry axis
\cite{ich14}.  Below, we call this axis a perpendicular axis.  Thus, the
torus isomer can rotate about a perpendicular axis, although the
collective rotation about the symmetry axis is quantum-mechanically
forbidden.  This rotational degree of freedom causes the precession
motion of the system as a whole.  Then, an interesting question arises
how such a ``femto-scale magnet'' rotates collectively to restore the
broken symmetry about a perpendicular axis.

A physical quantity characterizing such a collective rotation is the
moment of inertia about a perpendicular axis.  It has been theoretically
recognized that an independent-particle configuration in a deformed
harmonic-oscillator potential rotates with the rigid-body moment of
inertia when the self-consistency between the mean-field potential and
the density is fulfilled~\cite{BMbook}.  However, measured moments of
inertia for the case of the precession motion of prolately deformed
nuclei are often much smaller than the rigid-body values even when
pairing correlations are negligible~\cite{del04,shi05}.  This is because
of shell effects in high-$K$ prolate isomers \cite{del04}.  Although
precession modes of high-$K$ oblate isomers have not been observed yet,
their moments of inertia would be much reduced from the rigid-body
values due to oblate shell structures at small
deformations~\cite{and81}.

From these observations, it might be conjectured that the moment of
inertia about a perpendicular axis for the torus isomer also
significantly deviates from the classical rigid-body value, because the
torus isomer is a unique quantum object characterized by the alignment
of angular momenta of independent-particle motions.  It is thus
surprising that the moment of inertia about a perpendicular axis,
evaluated with the time-dependent Hatree-Fock (TDHF) method, from the
rotational period of the precession motion of the torus isomer in
$^{40}$Ca takes a value close to the classical rigid-body
value~\cite{ich14}.  We analyzed the microscopic structure of the
precession motion by using the random-phase approximation (RPA) method.
In the RPA calculation, the precession motion of the torus isomer is
generated by a coherent superposition of many one-particle-one-hole
excitations across the sloping Fermi surface. We found that the
precession motion obtained by the TDHF calculation is a pure collective
motion well decoupled from other collective modes. In our previous
studies, however, we focused only on the torus isomer of $^{40}$Ca.  It
is thus important to investigate whether torus isomers exist also in
other nuclei and the properties of the precession motion found there are
universal or not.

In this paper, we first perform a systematic investigation of the
high-spin torus isomers for a series of $N=Z$ even-even nuclei from
$^{28}$Si to $^{56}$Ni.  We show that the high-spin torus isomer of
$^{40}$Ca is not an exceptional case.  About forty years ago, Wong
suggested, using a macroscopic-microscopic method, the possible
existence of torus isomers at highly excited states of a wide region of
nuclei \cite{Wong73}.  Quite recently, Staszczak and Wong systematically
explored the existence of torus isomers using the constrained cranked
Hartree-Fock (HF) method and found some torus isomers at highly excited
states in several nuclei~\cite{Stas14}.  However, they use the
harmonic-oscillator basis expansion method, which is insufficient to
treat unbound states. It is therefore difficult to examine the stability
of the torus isomers against the nucleon emission in their calculation,
although some of them would contain single particles in unbound states.

We then perform a systematic TDHF calculation to investigate the
properties of the precession motion.  For all the high-spin torus
isomers obtained by the cranked HF calculation, we find the periodic
solutions of the TDHF equation of motion, which describe the precession
motions.  Among them, the precession motion of the 60 $\hbar$ torus
isomer of $^{40}$Ca is particularly stable and continues for many
periods.

To understand the microscopic origin of appearance of the torus isomers,
we analyze the process during which the shell structure of the large
oblate shape and that of the torus shape grow up from that of the
spherical shape.  Using the radially displaced harmonic-oscillator
(RDHO) model~\cite{Wong73} and the oblately deformed harmonic-oscillator
potential, we finally discuss why the lowest $0s$ components
disappear from all the single-particle wave functions of the occupied
states and how a large `hole' region is created in the center of the
nucleus to generate the torus shape.

This paper is organized as follows: In Section II, we describe the
theoretical framework and parameters of the numerical calculation.  In
Section III, we present results of the systematic calculation for static
and dynamical properties of the high-spin torus isomers including their
precession motions.  In Section IV, we analyze microscopic shell
structures of the torus isomers and discuss the reason why the torus
shape emerges beyond the limit of large oblate deformation.  Finally, we
summarize our studies in Section V.

\section{ Theoretical framework}
\subsection{Cranked HF calculation}
To investigate systematically the existence of high-spin torus isomer
states in a wide range of nuclei, we use the cranked three-dimensional
Skyrme HF method. To build high-spin states rotating about the symmetry
axis of the density distribution ($z$ axis), we add a Lagrange
multiplier, $\omega$, to the HF Hamiltonian, $\hat{H}$. Then, the
effective HF Hamiltonian, $\hat{H}'$, is written as
$\hat{H}'=\hat{H}-\omega\hat{J_z}$, where $\hat{J_z}$ denotes the $z$
component of the total angular momentum. We minimize this effective HF
Hamiltonian with a given Lagrange multiplier, which is equivalent to the
cranked HF equation given by
$\delta\left<\hat{H}-\omega\hat{J_z}\right>=0$.

For this purpose, we slightly modify the code {\tt Sky3d}. The details
of the code are given in Ref.~\cite{sky3d}. In the code, the
single-particle wave functions are described on a Cartesian grid with a
grid spacing of 1.0 fm, which is a good approximation for not only bound
states but also unbound states in contrast to the harmonic-oscillator
basis expansion.  We take $32\times32\times24$ grid points for the $x$,
$y$, and $z$ directions, respectively. This is sufficiently accurate to
provide converged configurations. The damped-gradient iteration method
\cite{gradient} is used, and all derivatives are calculated with the
Fourier transformation method.

In the calculation, we use the SLy6, SkI3, and SkM$^*$ Skyrme forces to
check the interaction dependence of the calculated results. These
effective interactions were well constructed based on nuclear bulk
properties but differ in details; SLy6 as a fit which includes
information on isotopic trends and neutron matter~\cite{Cha97a}, SkI3 as
a fit taking into account the relativistic isovector properties
of the spin-orbit force~\cite{Rei95a}, and SkM$^*$ as a widely used
traditional standard~\cite{Bar82}. However, except for the effective
mass, the bulk properties (equilibrium energy and density,
incompressibility, and symmetry energy) are comparable to each other.
In the energy density functional, we omit terms depending on the spin
density, because it may be necessary to extend the standard form of the
Skyrme interaction in order to properly take into account the
spin-density dependent effects \cite{todd_a} (see also a review
\cite{todd_b}), but such effects are inessential to the torus isomers.

\subsection{Setting of initial configurations}
In the cranked HF calculations, we first search for stable torus
configurations in a series of $N=Z$ even-even nuclei from $^{28}$Si to
$^{56}$ Ni. We use, as an initial configuration of the HF calculation,
an $\alpha$-cluster ring configuration placed on the $x$-$y$ plane, as
shown in Fig.~1 of Ref.~\cite{ich12}.  The $\alpha$-cluster wave
function is described by a Gaussian function with the width of 1.8 fm.
The center positions of the Gaussian functions are placed equiangularly
along a circle with a radius of 6.5 fm on the $(x,y)$ plane.  Only for
the calculations of $^{52}$Fe with the SkM$^{*}$ interaction, we use a
radius of 7.55 fm and a width of 1.63 fm.  Using these initial
configurations, we perform 15,000 HF iterations. We search for stable
torus solutions varying $\omega$ from 0.5 to 2.5 MeV/$\hbar$ with a step
of 0.1 MeV/$\hbar$.  After these calculations, we check the convergence
of the total energies, the density distributions, and the total angular
momenta.  In the calculations of the excitation energies, we subtract
the expectation value of the center-of-mass motion in both the ground
and the torus isomer states.

We next calculate all single-particle states including those above the
Fermi energy. To calculate those, we use, as initial wave functions of
the HF calculation, the single-particle wave functions of the RDHO
model~\cite{Wong73}.  This model is a good approximation to the
mean-field of torus-shaped isomers.  In this model, the single-particle
potential is given by
\begin{equation}
 V(r,z)=\frac{1}{2}m\omega^2_0(r-R_0)^2+\frac{1}{2}m\omega^2_0z^2,
\end{equation}
where $m$ denotes the nucleon mass, $\omega_0$ the oscillator frequency,
$r$ and $z$ the radial and the $z$ components of the cylindrical
coordinate system, and $R_0$ the radius parameter of the torus shape.
Since the radial wave function of the lowest energy in the RDHO model is
described by a shifted Gaussian function with the width
$d=\sqrt{\hbar/m\omega_0}$, we determine $\omega_0$ from the radius of a
cross section of a torus ring.  The optimal values of $R_0$ and $d$ are
determined through the global investigation mentioned above.  Using this
initial condition and a value of $\omega_0$ obtained by the global
investigation, we perform the HF iteration over 20000 times and
calculate the single-particle states up to the 40th for both protons and
neutrons.

\subsection{Sloping Fermi surface}
It is important to note that the cranking term $-\omega\hat{J_z}$ does
not change the single-particle wave functions for rotation about the
symmetry axis. Thus, it is useful to introduce the concept of
``sloping'' Fermi surface.  As usual, the single-particle Hamiltonian is
given by $\hat{H'}=\sum_i (\hat{h_i}-\omega\hat{j_z}^{(i)})$, where
$\hat{h_i}$ and $\hat{j_z}^{(i)}$ denote the mean-field Hamiltonian and
the $z$ component of the total angular momentum for each single
particle, respectively.  The eigenvalue of $\hat{H'}$ is written as
$E'=\sum_i[(e_i- \lambda)-\hbar\omega\Omega_i]$, where $\lambda$ denotes
the Fermi energy at $\omega=0$. The symbols $e_i$ and $\Omega_i$ denote
the single-particle energy and the eigenvalue of $\hat{j_z}^{(i)}$,
respectively.  By introducing the sloping Fermi surface defined by
$\lambda'(\Omega)=\lambda+\hbar\omega\Omega$, we can rewrite $E'$ as
$E'=\sum_i\{e_i-\lambda'(\Omega_i)\}$.  Therefore, aligned
configurations can be easily constructed by plotting the single-particle
energies as a function of $\Omega$ and tilting the Fermi surface in the
$(e, \Omega)$ plane.  It is important to note that the value of $\omega$
to specify an aligned configuration is not unique.  As we can
immediately see in Figs. 3-7 below, individual configurations do not
change for a finite range of $\omega$.

\subsection{Optimally aligned torus configurations}
Let us focus on optimally aligned torus configurations where all the
single-particle states below the sloping Fermi surface are occupied.
They are expected to be more stable than other aligned configurations
involving particle-hole excitations across the sloping Fermi surface.
Before carrying out the cranked HF calculations, we can easily presume
candidates of optimally aligned torus configurations.  Since the effects
of the spin-orbit potential are negligibly weak in the torus
configurations, not only $\Omega$ but also the $z$ component of the
orbital angular momentum, $\Lambda$, are good quantum numbers
($\Omega=\Lambda+\Sigma$, where $\Sigma$ denotes the $z$ component of
the spin, $\pm1/2$)~\cite{ich12}. Single-particle states having the same
$\Lambda$ value with different spin directions are approximately
degenerated and simultaneously occupied .  Thus, the lowest-energy
configurations for the torus shapes at $\omega=0$ are $\Lambda=0$,
$\pm1$, $\pm2$, and $\pm3$ for $^{28}$Si, $\Lambda=0$, $\pm1$, $\pm2$,
$\pm3$, and $+4$ or $-4$ for $^{32}$S, $\Lambda=0$, $\pm1$, $\cdots$,
$\pm4$ for $^{36}$Ar, $\Lambda=0$, $\pm1$, $\cdots$, $\pm4$, and $+5$ or
$-5$ for $^{40}$Ca, $\Lambda=0$, $\pm1$, $\cdots$, $\pm5$ for $^{44}$Ti,
$\Lambda=0$, $\pm1$, $\cdots$, $\pm 5$, and $+6$ or $-6$ for $^{48}$Cr,
$\Lambda=0$, $\pm1$, $\cdots$, $\pm6$ for $^{52}$Fe, and $\Lambda=0$,
$\pm1$, $\cdots$, $\pm6$, and $+7$ or $-7$ for $^{56}$Ni.

For instance, in $^{40}$Ca, possible aligned configurations at
$\omega\ne0$ are (i) $\Lambda=0$, $\pm1$, $\cdots$, $\pm4$, and $+5$ for
$J_z=20\hbar$ [$=5\hbar\times2$ (spin degeneracy) $\times2$ (isospin
degeneracy)], (ii) $\Lambda=0$, $\pm1$, $\pm2$, $\pm3$, $+4$, and $+5$
for $J_z=60\hbar$ [$=15\hbar\times2\times2$], and (iii) $\Lambda=0$,
$\pm1$, $\pm2$, $+3$, $+4$, $+5$, $+6$, and $+7$ for $J_z=100\hbar$
[$=25\hbar\times2\times2$].  However, we could not obtain stable HF
solutions for the configurations (i) and (iii): the centrifugal force is
insufficient for stabilizing the configuration (i), while the last
occupied single-particle state with $\Lambda=7$ is unbound for the
configuration (iii).  Indeed, we confirmed that the torus isomer
configuration (iii) with $J_z=100~\hbar$ slowly decays.  In the
systematic calculations, it is often difficult to discuss the stability
of torus isomers when such unbound states are included. To avoid this
difficulty, in this paper, we focus on torus configurations without
involving unbound single-particle states.

\subsection{TDHF calculation for the precession motion}

For the stable torus isomers obtained above, we performed TDHF
calculations to investigate their precession motions.  The time
evolution of the density distribution is determined by solving the TDHF
equation of motion $i\hbar\dot{\rho}=[\hat{H},\rho]$.  When an impulsive
force is provided in a direction perpendicular to the symmetry axis at
$t=0$, the torus isomer starts to execute the precession motion. This
precession motion is associated with a rotation about a perpendicular
axis, i.e., an axis perpendicular to the symmetry axis.  In
Ref.~\cite{ich14}, we already showed that this precession motion is a
pure collective motion to restore the broken symmetry and well described
as coherent superpositions of many 1p-1h excitations across the sloping
Fermi surface.  We investigate whether other torus isomers also execute
the precession motion well decoupled from other collective modes and
whether their moments of inertia are close to the rigid-body values or
not.  In this way, we can also check the stability of the obtained torus
isomers against given impulsive forces.

\begin{figure}[t]
\includegraphics[keepaspectratio,width=5cm]{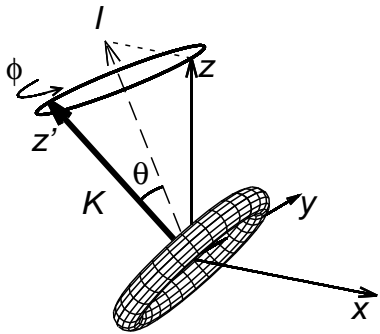}
\caption{Schematic picture for the precession motion of torus isomers
taken from Ref.~\cite{ich14}. The bold solid line denotes the symmetry
axis of the density distribution. The dashed line denotes the precession
axis. The symbols $\theta$ and $\phi$ denote the tilting and the
rotational angles, respectively.}  \label{pic_prec}
\end{figure}

Figure \ref{pic_prec} illustrates the schematic picture of the
precession motion.  At $t=0$, the torus isomer is placed on the $x$-$y$
plane with the angular momentum $K$ ($=J_z$) along the $z$ axis in the
laboratory frame.  When an impulsive force is provided in the negative
$x$ direction (the dotted line) at $t=0$, the total angular momentum
becomes $\vec{I}$ (the dashed line).  We call this vector the precession
axis.  After that, the symmetry axis of the density distribution in the
body-fixed frame (the bold solid line) starts to rotate about the
precession axis with the rotational angle $\phi$.  In the precession
motion, the value $K$ is conserved and its direction is identical to the
bold solid line.  The tilting angle $\theta$ is defined as the angle
between the bold solid and the dashed lines.  Then, the moment of
inertia for the rotation about a perpendicular axis,
$\mathscr{T}_\perp$, can be estimated by
$\mathscr{T}_\perp=I/\omega_{\rm prec}$, where $\omega_{\rm prec}$
denotes the rotational frequency of the precession motion.  To build the
first excited state of the precession motion, we provide an impulsive
force such that the total angular momentum becomes $I=K+1$.
  
To solve the TDHF equation, we use the code {\tt Sky3d} and take the
Taylor expansion of the time-development operator up to the 12th order.
The setups of spatial grid points and interactions are the same as those
of the cranked HF calculations described above. We start to perform
calculations from the initial density distribution obtained by the
cranked HF calculations. The time step of the TDHF calculations is 0.2
fm/$c$. We calculate the time evolution until 3000 fm/$c$.  To excite
the precession motion, we provide an impulsive force at $t=0$ by the
external potential given by $V_{\rm ext}(r,\varphi,z)=V_0 z \cos \varphi
\exp[-(r-R_0)^2/d^2]$.  This impulsive force gives an angular momentum
in the negative $x$ direction at $t=0$.  The parameter $V_0$ is chosen
such that the total angular momentum becomes $I=K+1$.

\begin{table}[t]
 \caption{\label{tab1} Stable torus isomers obtained in the cranked HF
 calculation with various Skyrme interactions. The excitation energy,
 $E_{\rm ex}$, is measured from the ground state.  The calculated
 density distributions are fitted to the Gaussian function
 $\rho(r,z)=\rho_0 e^{-\left[(r-R_0)^2+z^2\right]/d^2}$ and the
 resulting values of the parameters, $\rho_0$, $R_0$, and $d$, are
 listed.  The symbols, $\mathscr{T}_\perp^{\rm rid}$ and
 $\mathscr{T}_\parallel^{\rm rid}$, denote the rigid-body moments of
 inertia for the rotations about a perpendicular and the symmetry axes,
 respectively. }
\begin{ruledtabular}
\begin{tabular}{cccccccc}
System & $J_z$ &$E_{\rm ex}$ &$\rho_0$ & $R_0$
 & $d$ &$\mathscr{T}_\perp^{\rm rid}$&$\mathscr{T}_\parallel^{\rm rid}$\\
&($\hbar$)&(MeV)&(fm$^{-3}$)&(fm)&(fm) &($\hbar^2$/MeV)&($\hbar^2$/MeV) \\
 \hline
 (SLy6)& & &\\
 $^{36}$Ar &36&123.89 &0.137 &5.12 & 1.62&14.3 &26.4 \\
 $^{40}$Ca &60&169.71 &0.129 &6.07 & 1.61&21.0 &39.6 \\
 $^{44}$Ti &44&151.57 &0.137 &6.30 & 1.61&24.6 &46.5 \\
 $^{48}$Cr &72&191.25 &0.132 &7.19 & 1.60&33.8 &64.7 \\
 $^{52}$Fe &52&183.70 &0.138 &7.47 & 1.60&39.1 &75.1 \\
 (SkI3)& & &\\
 $^{36}$Ar &36&125.15 &0.146 &5.01 &1.58 &13.7 &25.3\\
 $^{40}$Ca &60&173.52 &0.138 &5.90 &1.58 &19.9 &37.5\\
 $^{44}$Ti &44&153.02 &0.146 &6.17 &1.58 &23.6 &44.6\\
 $^{48}$Cr &72&193.66 &0.141 &7.00 &1.57 &32.0 &61.3\\
 $^{52}$Fe &52&183.70 &0.147 &7.31 &1.57 &37.5 &71.9\\
 (SkM$^*$)& & &\\ 
 $^{36}$Ar &36&124.80 &0.131 &5.16 &1.65 &14.6 &26.9\\
 $^{40}$Ca &60&167.84 &0.122 &6.17 &1.64 &21.8 &41.0\\
 $^{44}$Ti &44&152.20 &0.131 &6.36 &1.64 &25.1 &47.5\\
 $^{48}$Cr &72&192.40 &0.125 &7.30 &1.63 &34.9 &66.7\\
 $^{52}$Fe &52&187.08 &0.132 &7.55 &1.63 &40.0 &76.7\\
\end{tabular}
\end{ruledtabular}
\end{table}
  
\section{Results of calculation}

\begin{figure}[htbp]
\includegraphics[keepaspectratio,width=\linewidth]{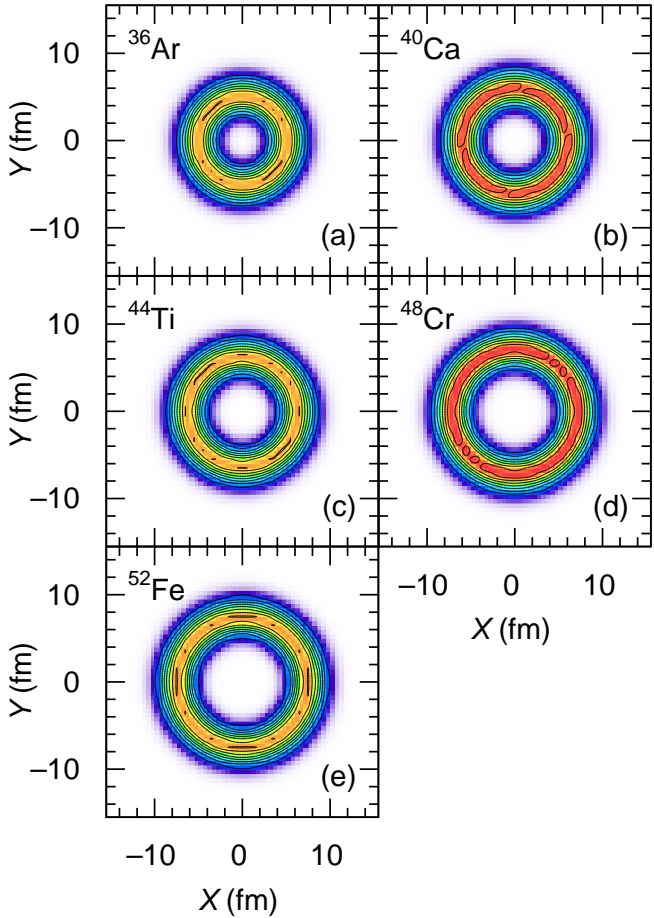}
\caption{(Color online) Density distributions on the $z=0$ plane of the
obtained stable torus isomers. The contours correspond to multiple steps
of 0.015 fm$^{-3}$. The color is normalized by the largest density in
each plot.}
\end{figure}

\begin{figure}[t]
\includegraphics[keepaspectratio,width=\linewidth]{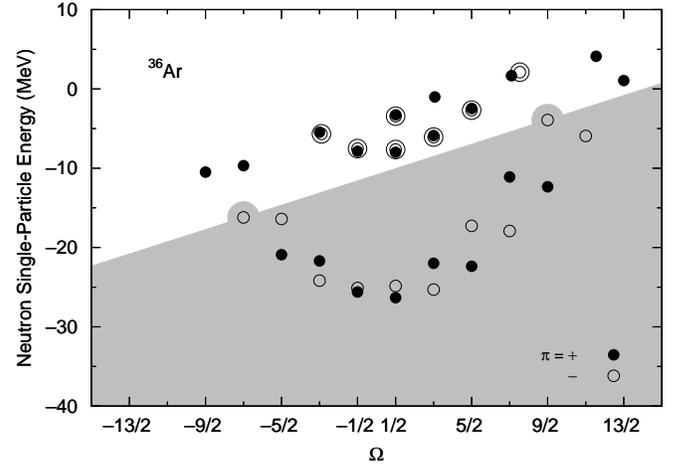}
\caption{Single-particle energies versus the $z$
component of the total angular momentum, $\Omega$, for $^{36}$Ar.  
Solid and open circles denote the single-particle energies of the
positive- and negative-parity states, respectively. To illustrate the
degeneracy of positive- and negative-parity states, some
negative-parity states are shown by  double open circles.}  \label{figsp1}
\end{figure}
\begin{figure}[t]
\includegraphics[keepaspectratio,width=\linewidth]{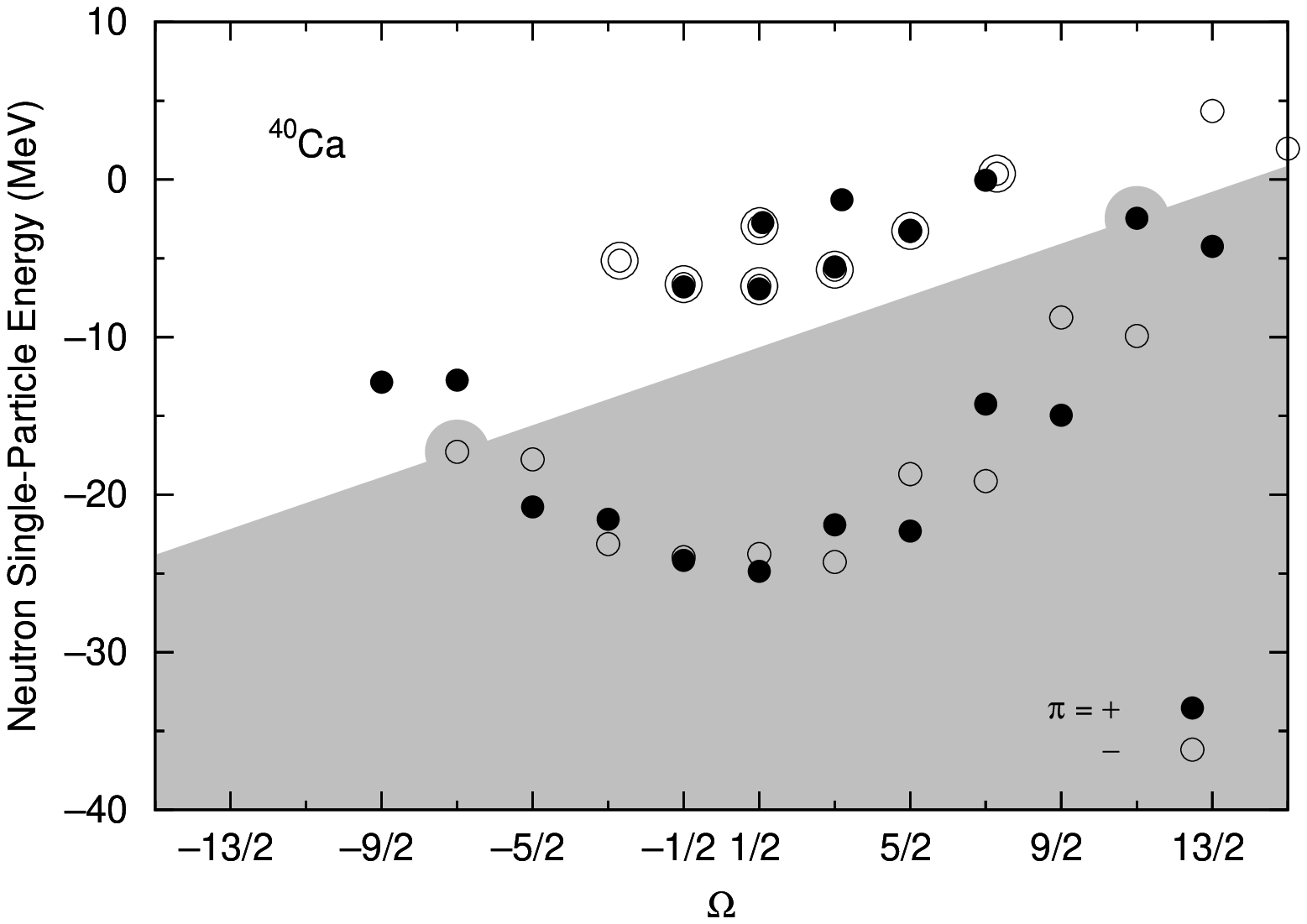}
\caption{Single-particle energies versus $\Omega$ for $^{40}$Ca.  All
symbols are the same as in Fig.~\ref{figsp1}.}  \label{figsp2}
\end{figure}

\begin{figure}[t]
\includegraphics[keepaspectratio,width=\linewidth]{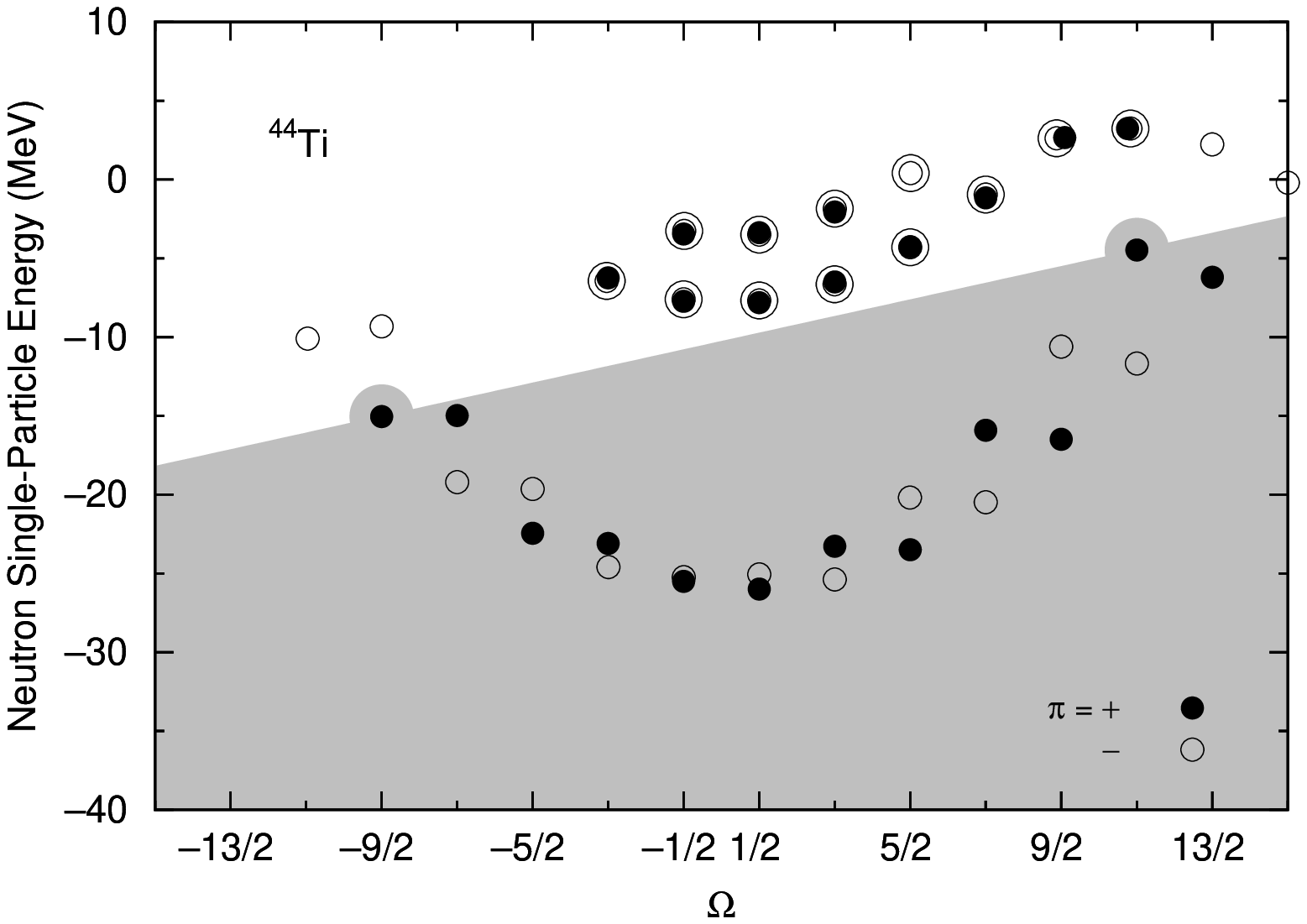}
\caption{Single-particle energies versus $\Omega$ for $^{44}$Ti.  All
symbols are the same as in Fig.~\ref{figsp1}.}  \label{figsp3}
\end{figure}

\begin{figure}[t]
\includegraphics[keepaspectratio,width=\linewidth]{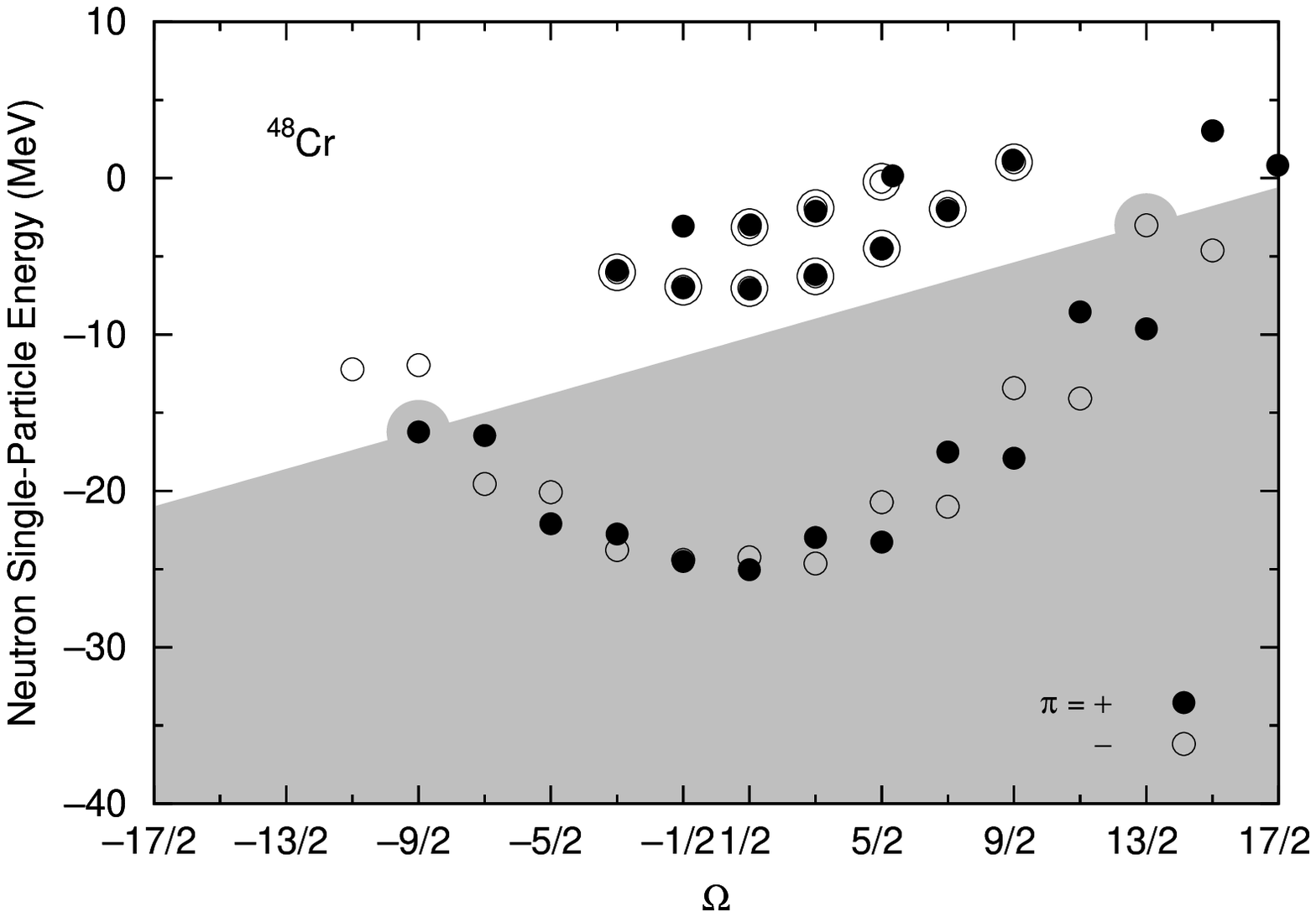}
\caption{Single-particle energies versus $\Omega$ for $^{48}$Cr.  All
symbols are the same as in Fig.~\ref{figsp1}.}  \label{figsp4}
\end{figure}
  
\begin{figure}[t]
\includegraphics[keepaspectratio,width=\linewidth]{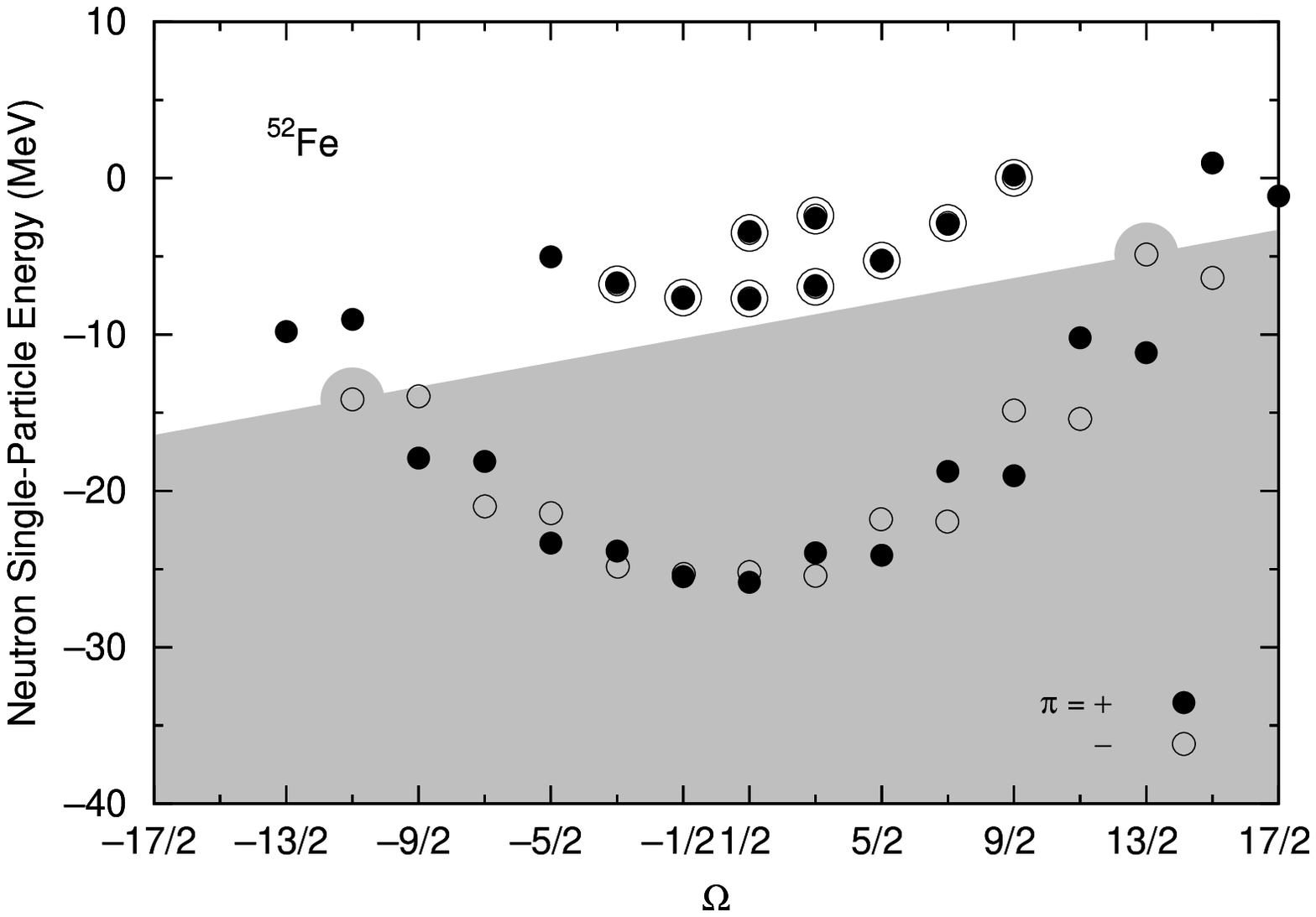}
\caption{Single-particle energies iversus $\Omega$ for $^{52}$Fe.  All
symbols are the same as in Fig.~\ref{figsp1}.}  \label{figsp5}
\end{figure}

\subsection{Static properties}

We have carried out a systematic search for stable torus isomers for the
$N=Z$ even-even nuclei from $^{28}$Si to $^{56}$Ni.  The result of the
calculation is summarized in Table \ref{tab1}.  We obtain the stable
torus isomers in $^{36}$Ar for $J_z=36$ $\hbar$, $^{40}$Ca for $J_z=60$
$\hbar$, $^{44}$Ti for $J_z=44$ $\hbar$, $^{48}$Cr for $J_z=72$ $\hbar$,
and $^{52}$Fe for $J_z=52$ $\hbar$ with all the three Skyrme
interactions used in this study.  On the other hand, we have not found
any stable torus isomer in $^{28}$Si, $^{32}$S, and $^{56}$Ni.  In
Fig.~\ref{figsp1}, we plot the total density distributions of the torus
isomers obtained in the cranked HF calculation with the SLy6
interaction.
  
The total density distribution of each of the torus isomers obtained in
the cranked HF calculation is well fitted by the Gaussian function
$\rho(r,z)=\rho_0e^{-\left[(r-R_0)^2+z^2\right]/d^2}$, where $\rho_0$,
$R_0$, and $d$ denote the maximum value of the nucleon density, the
radius of the torus ring, and the width of a cross section of the torus
ring, respectively.  The resulting values of the parameters, $\rho_0$,
$R_0$, and $d$, are tabulated in the middle part of Table \ref{tab1}.
We see that the values $\rho_0$ and $d$ are almost constant for all the
torus isomers. The interaction dependence of these values is weak. It is
interesting that, in all the results, $\rho_0$ is smaller than the
saturation nuclear density ($\rho_{\rm sat}\sim 0.17$ fm$^{-3}$) and $d$
is close to the width of an alpha particle used in Brink's
$\alpha$-cluster model ($d_\alpha \sim$ 1.46 fm) \cite{brink}.

Using the total density distribution, we also calculate the rigid-body
moments of inertia for rotation about a perpendicular axis,
$\mathscr{T}_\perp^{\rm rid}$, and the symmetry axis,
$\mathscr{T}_\parallel^{\rm rid}$.  The results are also shown in Table
\ref{tab1}.  Later, we shall compare these values for
$\mathscr{T}_\perp^{\rm rid}$ with those obtained by an analysis of the
precession motions in the TDHF time evolution.

To investigate microscopic structures of the torus isomers, we plot in
Figs.~\ref{figsp1}-\ref{figsp5} neutron single-particle energies versus
$\Omega$ for each torus isomer calculated with the SLy6 interaction.  In
the figures, the solid and the open circles denote the positive- and
negative-parity states, respectively. The gray area in each plot denotes
the occupied states.  In each plot, we see that the single-particle
energies with the same $\Lambda$ are almost degenerate. This indicates
that the effects of the spin-orbit force are negligibly small and
$\Lambda$ is approximately a good quantum number in all the torus
isomers.  One may also notice that the Kramer's degeneracy for a pair of
single-particle states with $\pm\Lambda$ is lifted.  This is due to the
time-odd components (dependent on the current density) of the cranked HF
mean fields associated with the macroscopic currents, which are produced
by the alignment of the single-particle angular momenta with large
values of $\Lambda$.

Because of the negligible spin-orbit splittings, the spin-orbit partners
are always occupied simultaneously.  Therefore, the $J_z$ values of the
optimally aligned configurations are easily determined by summing up the
$\Lambda$ values of the occupied single-particle states: they are
$\Lambda=0, \pm 1, \pm 2, \pm 3, +4, +5$ [$J_z$ = 9 $\hbar$ $\times 2$
(spin degeneracy) $\times 2$ (isospin degeneracy) $=36$ $\hbar$] for
$^{36}$Ar, $\Lambda=0, \pm 1, \pm 2, \pm 3, +4, +5, +6$ [$J_z$ = 15
$\hbar$ $\times 2 \times 2=60$ $\hbar$] for $^{40}$Ca, $\Lambda=0, \pm
1, \cdots, \pm 4, +5, +6$ [$J_z$ = 11 $\hbar$ $\times 2 \times 2
=44$ $\hbar$] for $^{44}$Ti, $\Lambda=0, \pm 1, \cdots, \pm 4, +5,
+6, +7$ [$J_z$ = 18 $\hbar$ $\times 2 \times 2 =72$ $\hbar$] for
$^{48}$Cr, and $\Lambda=0, \pm 1, \cdots, \pm 5, +6, +7$
[$J_z$ = 13 $\hbar$ $\times 2 \times 2 =52$ $\hbar$] for $^{52}$Fe.

\begin{figure}[t]
\includegraphics[keepaspectratio,width=\linewidth]{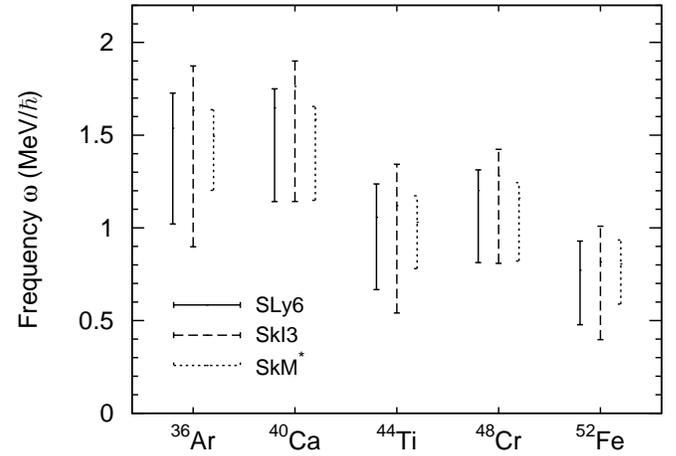}
\caption{Regions of $\omega$ for which each of the torus isomers
stably exists. The solid, dashed, and dotted lines denote the results 
calculated with the SLy6, SkI3, and SkM$^*$ interactions, respectively}
\label{region}
\end{figure}

We can estimate from the figures a region of $\omega$ for which each of
the torus isomers stably exists.  This is done by determining the
steepest and the most gradual slopes of the Fermi surface for which the
occupied single-particle configuration remains the same.  The results
are plotted in Fig.~\ref{region}.  The solid, dashed, and dotted lines
denote the regions of $\omega$ for each of the stable torus isomers
obtained with the SLy6, SkI3, and SkM$^*$ interactions, respectively.
We see that the result does not strongly depend on the Skyrme
interaction employed, although the width is weakly dependent on it.

\subsection{Dynamic properties}

\begin{table}[htbp]
 \caption{\label{tab2} Results of the TDHF calculation for the
 precession motions of the torus isomers from $^{36}$Ar to $^{52}$Fe.
 The symbol $I$ denotes the resulting total angular momentum after an
 impulsive force is provided.  The symbol $T_{\rm prec}$ denotes the
 average over the two periods from $t=0$ for the precession motion. The
 symbol $\omega_{\rm prec}$ denotes the precession frequency estimated
 by $\omega_{\rm prec}=2\pi/T_{\rm prec}$. The symbol
 $\mathscr{T}_\perp^{\rm TDHF}$ denotes the moment of inertia for the
 rotation about a perpendicular axis estimated by
 $\mathscr{T}_\perp^{\rm TDHF}=I/\omega_{\rm prec}$.}
\begin{ruledtabular}
\begin{tabular}{ccccc}
System & $I$ &$T_{\rm prec}$ &$\omega_{\rm prec}$ &
 $\mathscr{T}_\perp^{\rm TDHF}$\\
&($\hbar$)&(MeV/$\hbar$)&(MeV)&($\hbar^2$/MeV) \\
 \hline
 (SLy6)& & &\\
 $^{36}$Ar &37&450.1&2.75&13.5\\
 $^{40}$Ca &61&402.5&3.08&19.8\\
 $^{44}$Ti &45&651.0&1.90&23.7\\
 $^{48}$Cr &73&554.5&2.24&32.6\\
 $^{52}$Fe &53&872.8&1.42&37.3\\
 (SkI3)& & &\\
 $^{36}$Ar &37&427.9&2.90&12.8\\
 $^{40}$Ca &61&378.6&3.28&18.6\\
 $^{44}$Ti &45&624.4&1.99&22.7\\
 $^{48}$Cr &73&524.5&2.36&30.9\\
 $^{52}$Fe &53&839.0&1.48&35.9\\
 (SkM$^*$)& & \\
 $^{36}$Ar &37&464.2&2.67&13.9\\
 $^{40}$Ca &61&418.2&2.96&20.6\\
 $^{44}$Ti &45&666.1&1.86&24.2\\
 $^{48}$Cr &73&572.8&2.16&33.7\\
 $^{52}$Fe &53&894.8&1.39&38.3\\
\end{tabular}
\end{ruledtabular}
\label{precession}
\end{table}

\begin{figure*}[htbp]
 \begin{tabular}{ccc}

 \begin{minipage}{0.33\hsize}
 \includegraphics[keepaspectratio,width=\linewidth]{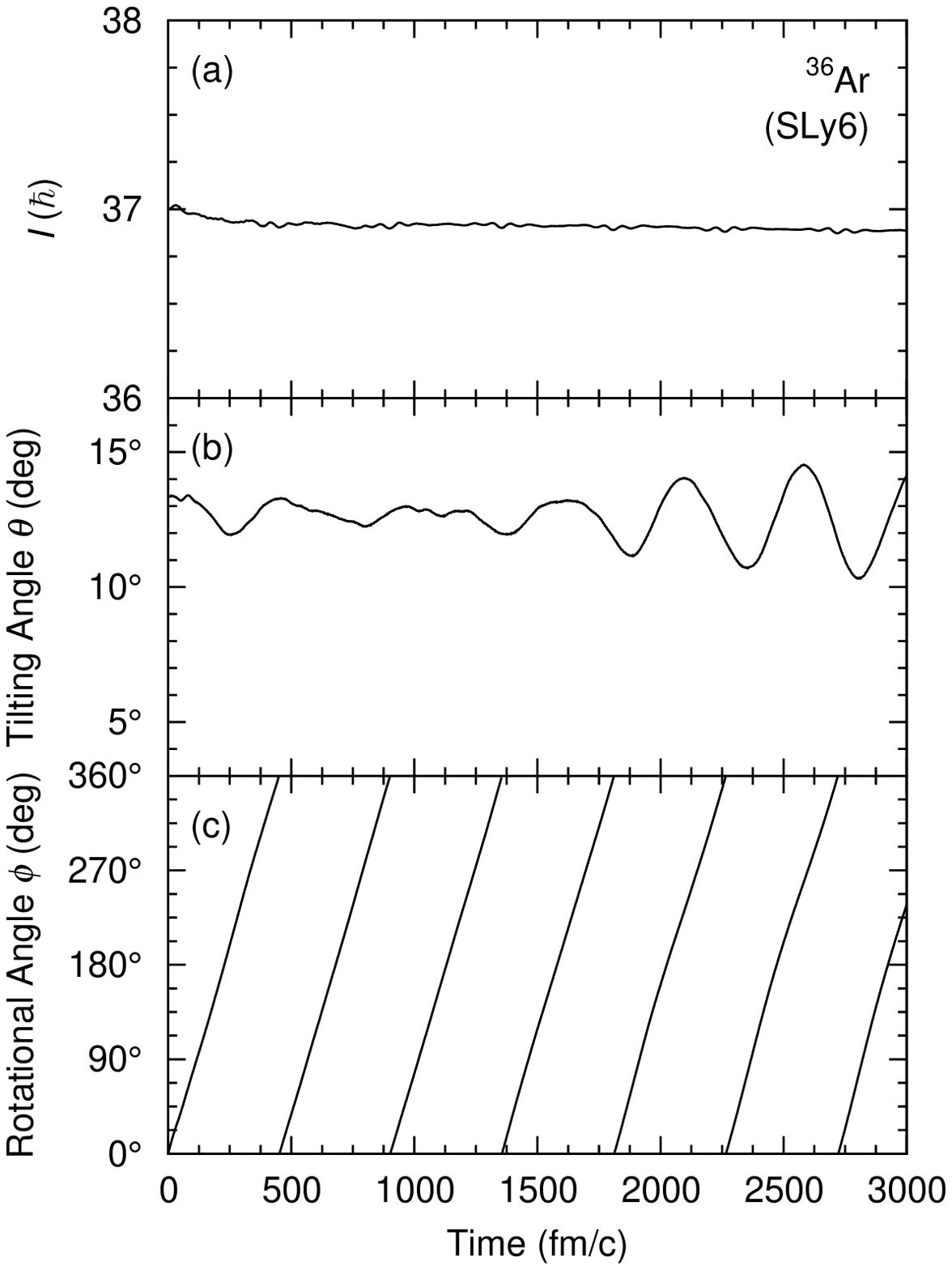}
 \end{minipage}
&
 \begin{minipage}{0.33\hsize}
 \includegraphics[keepaspectratio,width=\linewidth]{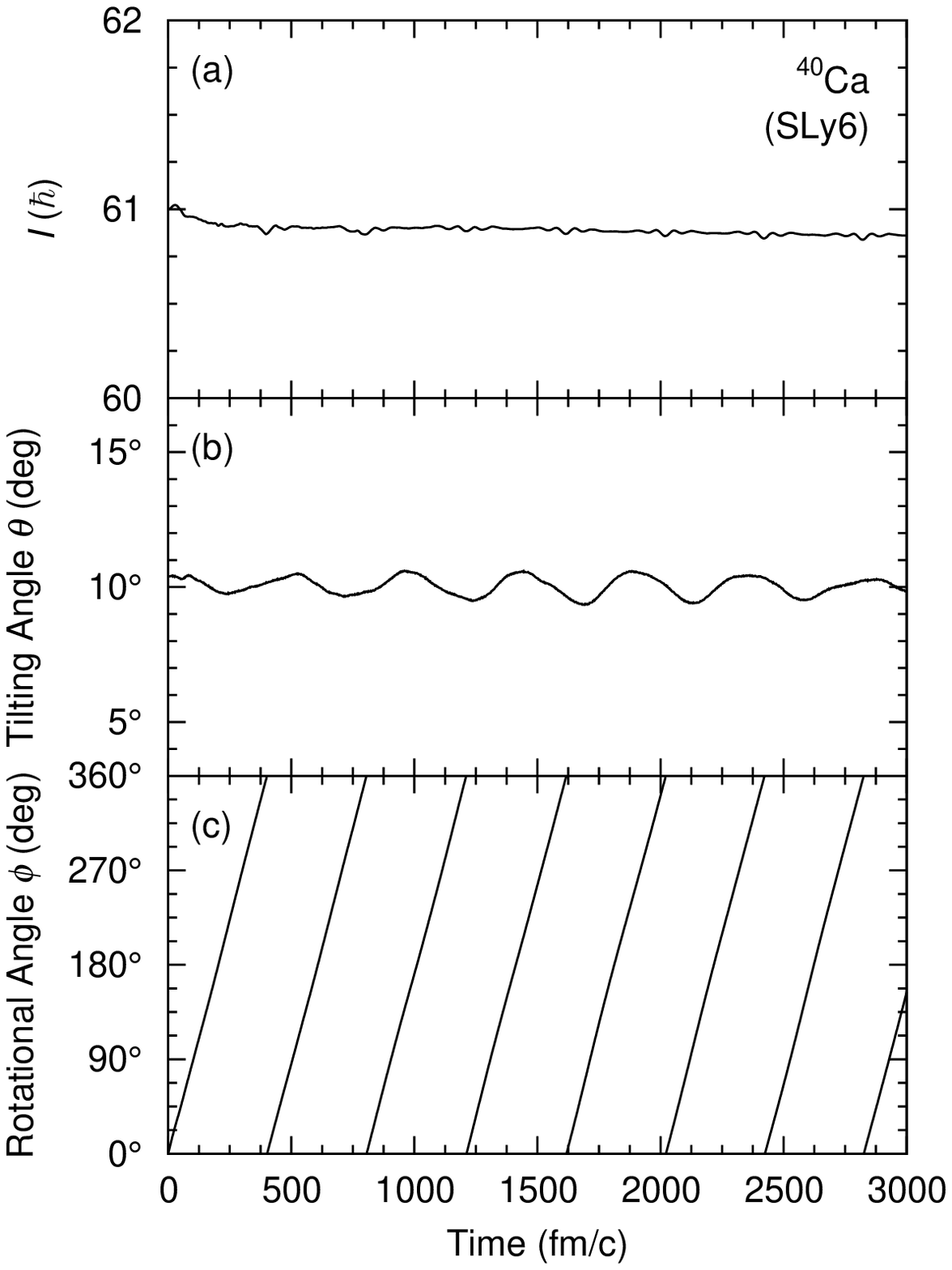}
 \end{minipage}
&
 \begin{minipage}{0.33\hsize}
 \includegraphics[keepaspectratio,width=\linewidth]{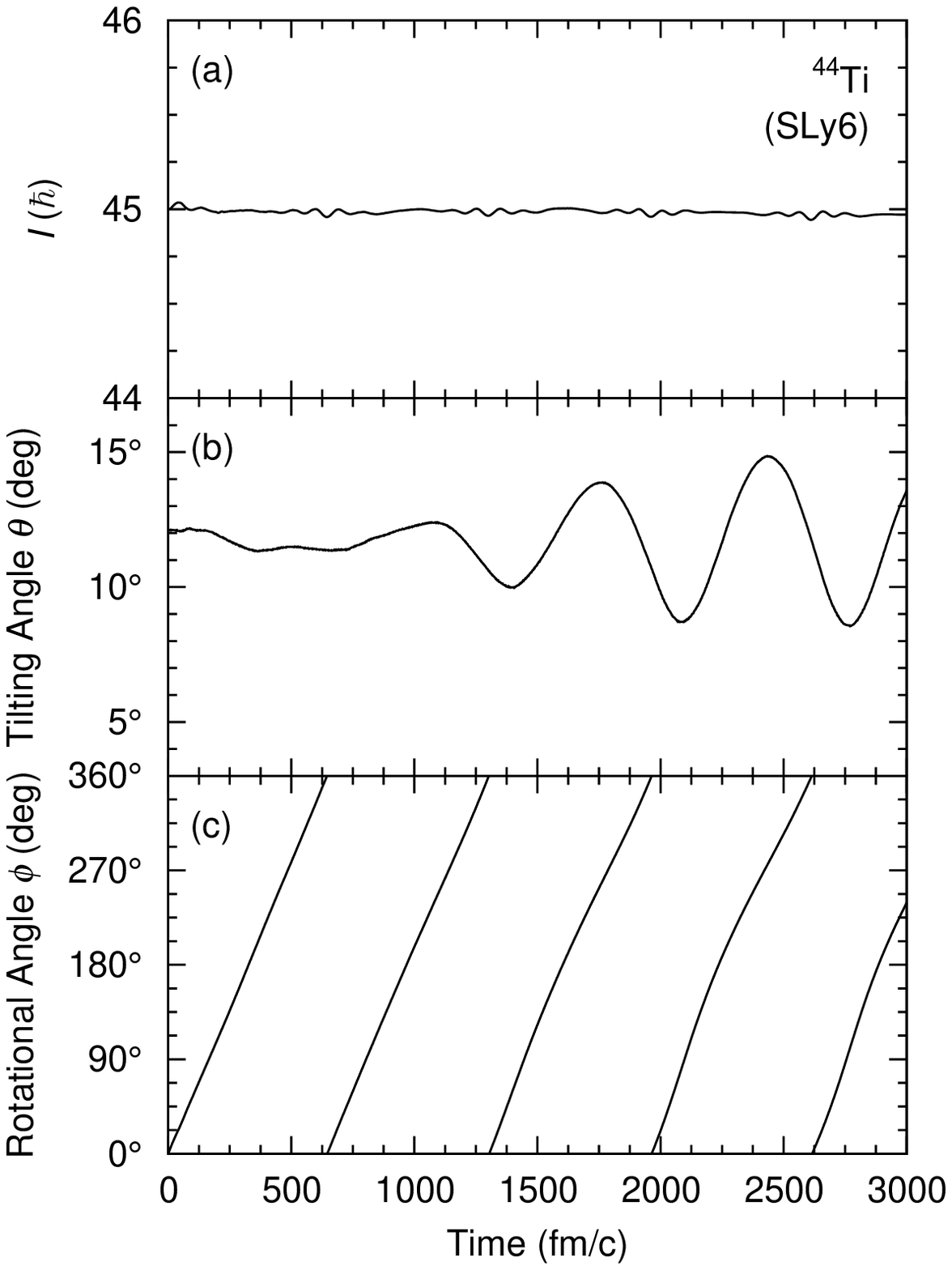}
 \end{minipage}
 \end{tabular}
 
\vspace{0.5cm}

 \begin{tabular}{cc}

  \begin{minipage}{0.33\hsize}
 \includegraphics[keepaspectratio,width=\linewidth]{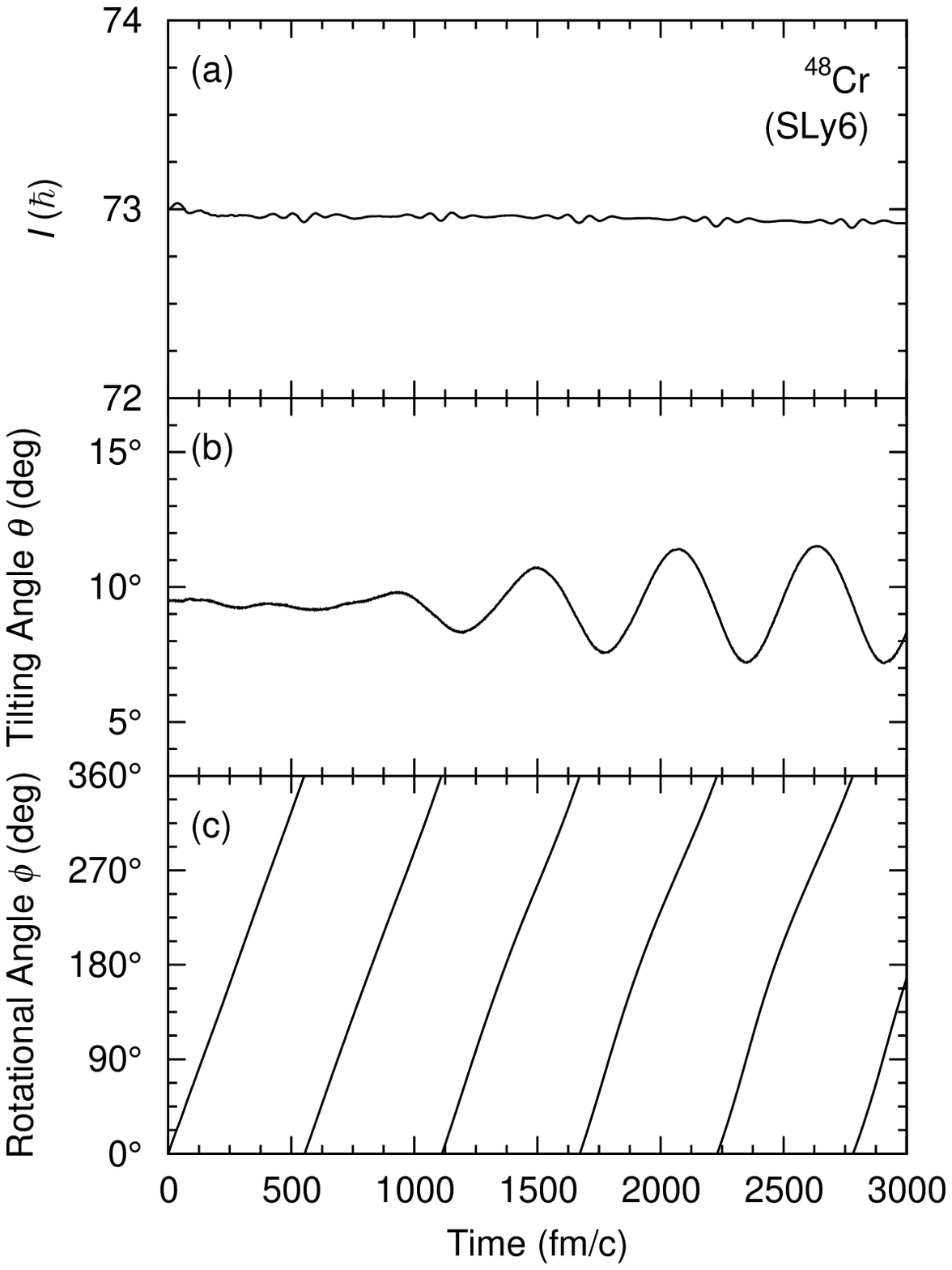}
  \end{minipage}
&
 \begin{minipage}{0.33\hsize}
 \includegraphics[keepaspectratio,width=\linewidth]{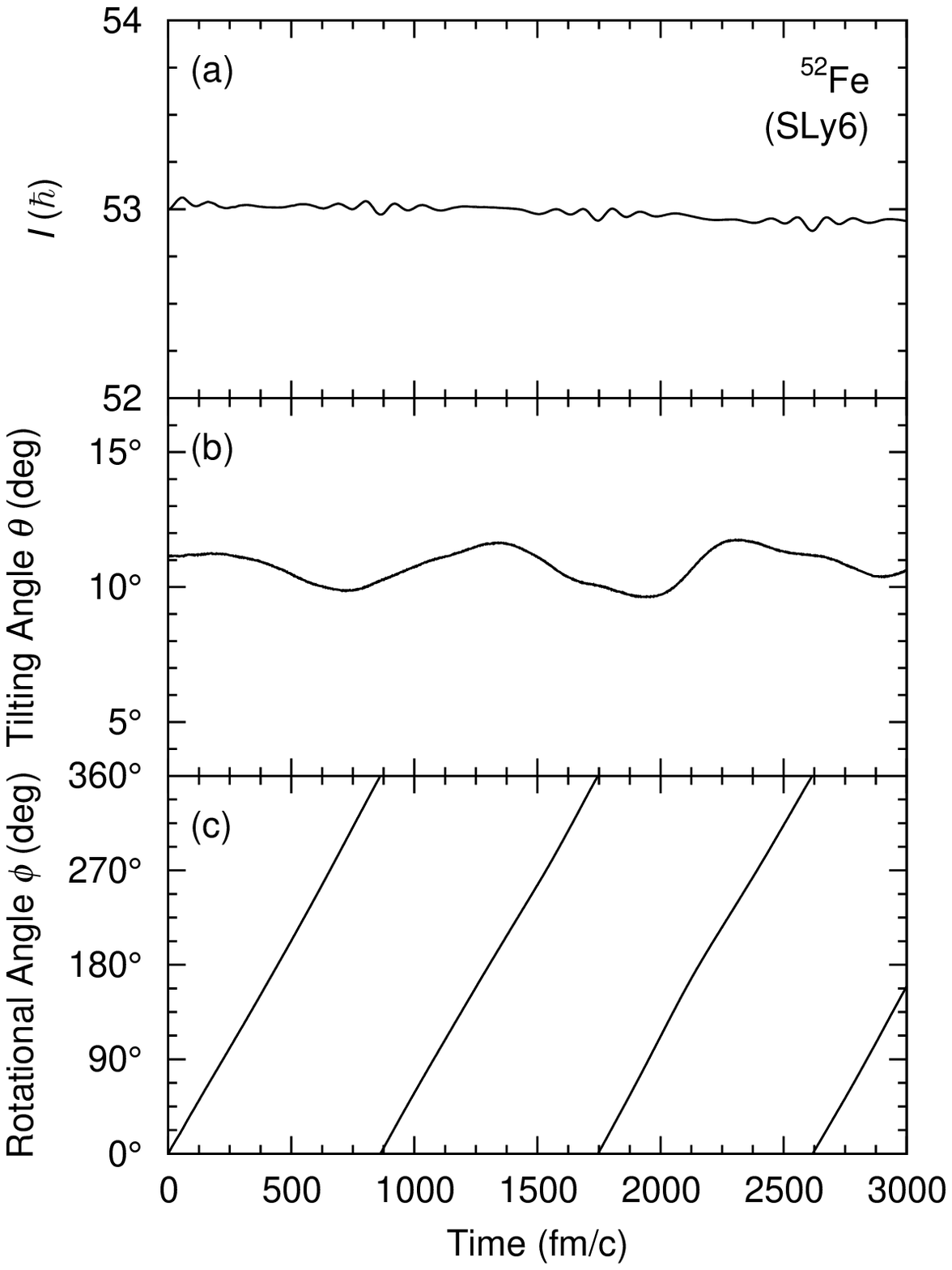}
 \end{minipage}

\end{tabular} 
\caption{Time evolution of the precession motion for each of
 the torus isomers from $^{36}$Ar to $^{52}$Fe 
 calculated by solving the TDHF equation of motion  
 for the SLy6 interaction. In each plot, panels (a), (b),
 and (c) denote the total angular momentum $I$, the tilting angle
 $\theta$, and the rotational angule $\phi$, respectively. } \label{tdhf}
\end{figure*}

We carried out a systematic TDHF calculation for each of the torus
isomers and found that that the TDHF time evolution of the density
distribution is quite similar to that displayed in Fig.~2 of
\cite{ich14}.  Figure \ref{tdhf} shows the calculated time evolution of
the precession motion for each of the torus isomers obtained with the
SLy6 interaction.  In each plot in Fig.~\ref{tdhf}, panels (a), (b), and
(c) denote the total angular momentum, $I$, the tilting angle, $\theta$,
and the rotational angle, $\phi$, respectively.  In panel (a) in each
plot, we can see that the total angular momentum is conserved very
well. This indicates that the TDHF calculations are sufficiently
accurate.  We find that the precession motion of the $^{40}$Ca torus
isomer is especially stable [see panel (b) in each plot], where the
rotational angle $\phi$ lineally increases with time, indicating that
the rotation of the symmetry axis about the precession axis keeps a
constant velocity through all the periods.  This indicates that the
strong shell effects responsible for the appearance of the torus isomer
in $^{40}$Ca also stabilize the precession motion.  We find that the
precession motions emerge also for other torus isomers and they are
stable at least for two periods.  After that, however, the tilting angle
gradually starts to fluctuate.  Correspondingly, the rotational angle
$\phi$ also starts to deviate from the linear time evolution [see panel
(c) in each plot].  We have also carried out similar TDHF calculations
with the use of the SkI3 and SkM$^*$ interactions.  The results are
similar to those shown above for the SLy6 interaction, which implies
that the properties of the precession motion are robust and depend on
the choice of the Skyrme interaction only weakly.

To evaluate the moment of inertia for the rotation about a perpendicular
axis, we take the average of the two periods starting from $t=0$ during
which the precession motion is especially stable.  The results are
tabulated in the third column of Table \ref{precession}.  Using these
values, we calculate the frequency of the precession motion by
$\omega_{\rm prec}=2\pi/T_{\rm prec}$ and the moment of inertia for the
rotation about a perpendicular axis by $\mathscr{T}_\perp^{\rm
TDHF}=I/\omega_{\rm prec}$.  The results are tabulated in the forth and
fifth columns of Table \ref{precession}.  The obtained moments of
inertia are very close to the rigid-body values tabulated in Table
\ref{tab1} for all the Skyrme interactions employed.  As discussed in
Ref.~\cite{ich14}, these results indicate that the precession motions
under consideration are pure collective motions generated by coherent
superpositions of many 1p-1h excitations across the sloping Fermi
surface.

\begin{figure}[t]
\includegraphics[keepaspectratio,width=\linewidth]{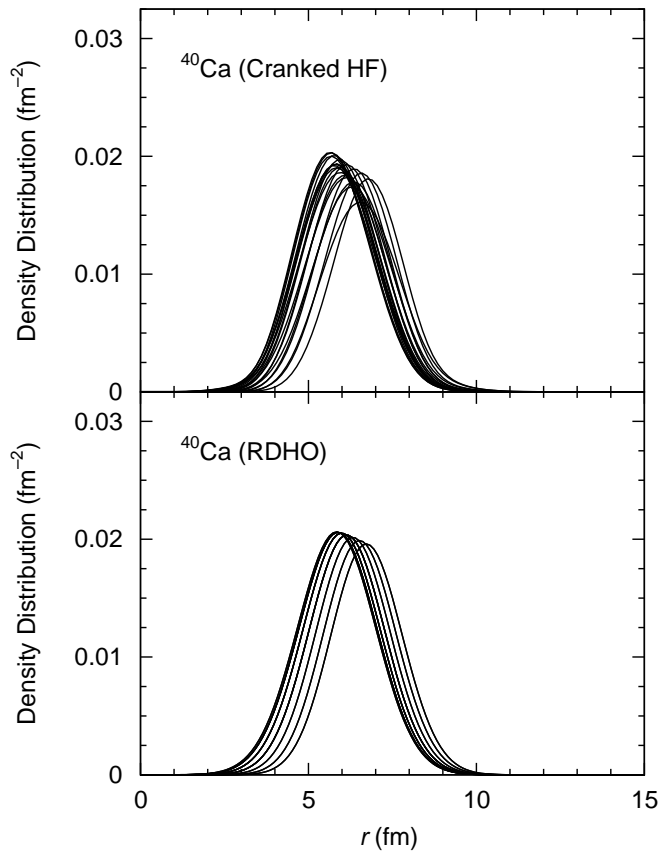}
\caption{Radial density distributions of individual single-particle
states on the $z=0$ plane for $^{40}$Ca obtained by the cranked HF
calculations (upper panel) and the RDHO model (lower pannel).  The
densities for the $\phi$ direction in cylindrical coordinates are
integrated.}  \label{wfrdho}
\end{figure}

\begin{figure}[http]
\includegraphics[keepaspectratio,width=\linewidth]{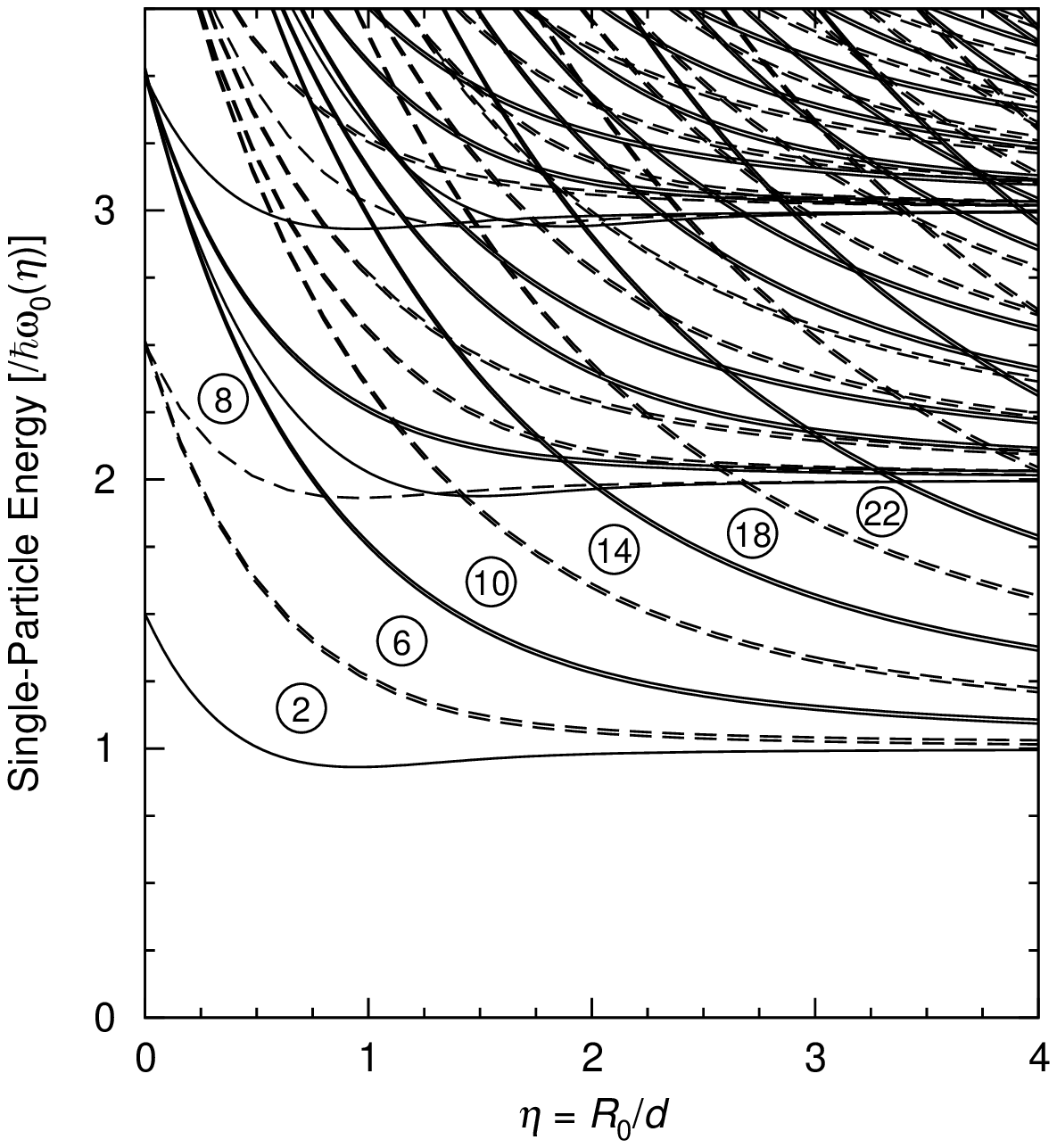}
\caption{Nilsson diagram versus $\eta=R_0/d$ of the RDHO model for
$^{40}$Ca. The solid and dashed lines denote the single-particle states
with positive- and negative-parities, respectively. The single-particle
energies are plotted in units of $\hbar\omega_0(\eta)$. To illustrate the
degeneracy of the levels, the single-particle energies with higher
$\Omega$ are slightly shifted.} \label{nilrdho}
\end{figure}

\section{Discussion}
\subsection{Radial density distributions of individual single-particle states}

Let us examine the radial density distributions of individual
single-particle wave functions on the $z=0$ plane for the torus isomer
of $^{40}$Ca.  For this purpose, we interpolate the density
distributions described with a Cartesian coordinate in the cranked HF
calculations by means of a third-order B-spline function.  After
that, we transform those to a cylindrical coordinate representation
[$\rho_i(x,y)\rightarrow \rho_i(r,\varphi)$]. We then integrate
$\rho_i(r,\varphi)$ in the $\varphi$ direction and obtain $\rho_i(r)$.
The calculated results are plotted in the upper panel of
Fig.~\ref{wfrdho}.

As shown in Refs.~\cite{ich12,ich14}, the RDHO model can describe well
the microscopic structures of torus isomers. To illustrate this, we
solve the Schr\"odinger equation with the RDHO potential, Eq.~(1), by
means of the deformed harmonic-oscillator basis expansion and calculate
$\rho_i(r)$. In the calculation, we take $R_0=6.07$ fm and $d=1.61$ fm
for the RDHO model and the same aligned single-particle configuration as
that obtained by the cranked HF calculation for $^{40}$Ca. The obtained
radial density distributions of the individual single-particle states
are plotted in the lower panel of Fig.~\ref{wfrdho}. Using these density
distributions, we calculate the rigid-body moments of inertia about a
perpendicular axis and the symmetry axis: they are
$\mathscr{T}_\perp^{\rm RDHO}=21.3$ $\hbar^2$/MeV and
$\mathscr{T}_\parallel^{\rm RDHO}=40.2$ $\hbar^2$/MeV, respectively.
These values are in good agreement with those obtained by the cranked HF
calculation.

In Fig.~\ref{wfrdho}, it is clearly seen that the radial density
distributions of the individual single-particle states in the RDHO model
are quite similar to those obtained by the cranked HF calculations.  In
particular, the peak positions of each radial density distribution are
in good agreement between the two calculations. As a matter of fact, the
peak position of each density distribution shifts to a larger $r$ with
increasing orbital angular momentum. Looking into details of the density
distributions obtained by the cranked HF calculations, one may notice
that some radial density distributions with high angular momentum
slightly shift due to the spin-orbit potential. In Fig.~\ref{figsp2},
the degeneracy of single-particle energies with the same high $\Lambda$
is indeed slightly broken for the spin-orbit partner with
$\Omega^\pi=9/2^-$ and $11/2^-$ ($\Lambda=5$) and that with
$\Omega^\pi=11/2^+$ and $13/2^+$ ($\Lambda=6$). These spin-orbit effects
are absent in the RDHO model.

\begin{figure*}[htbp]
  \includegraphics[keepaspectratio,width=\linewidth]{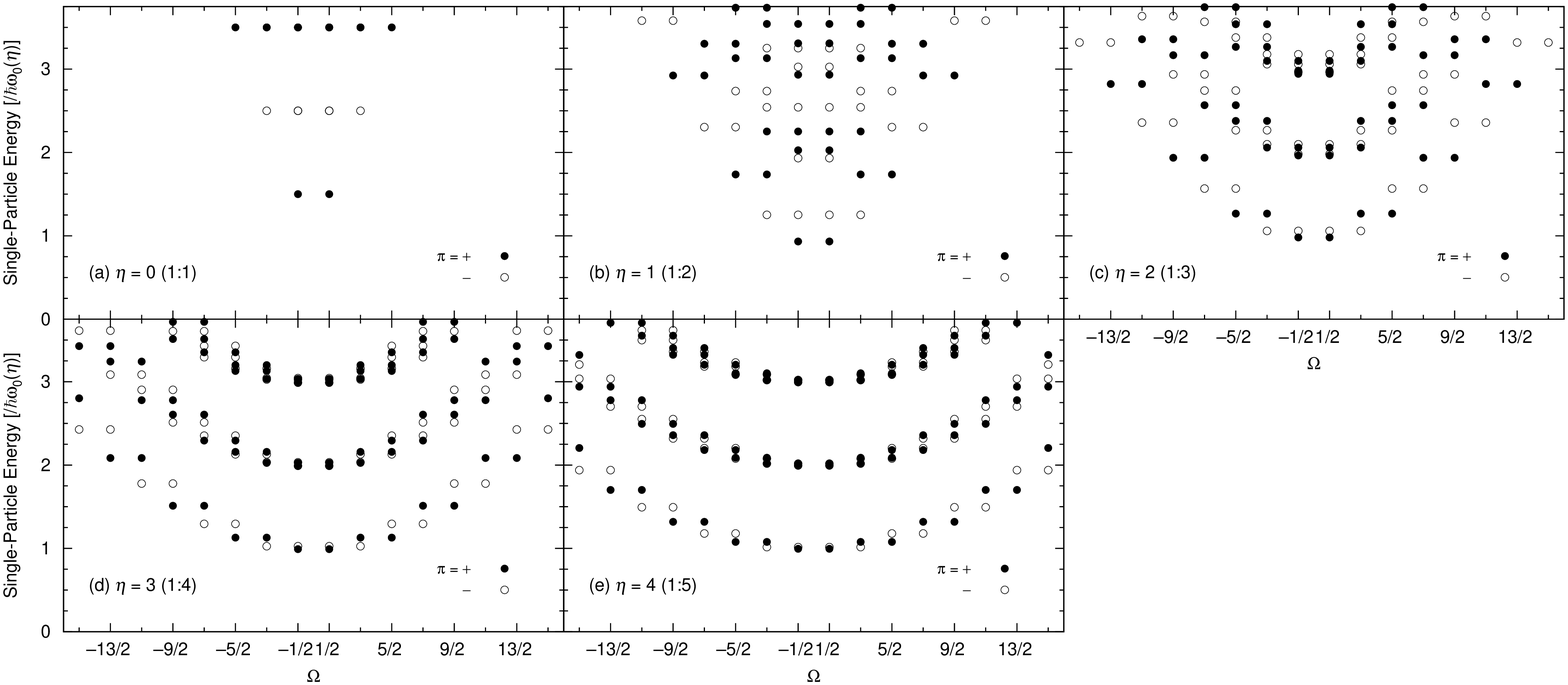}
  \caption{Single-particle energies of the RDHO model versus $\Omega$ at
  various values of $\eta=R_0/d$. The solid and open circles denote the
  single-particle states with positive and negative parities,
  respectively.} \label{spc_rdho}
 \end{figure*}
\begin{figure}[b]
\includegraphics[keepaspectratio,width=8.cm]{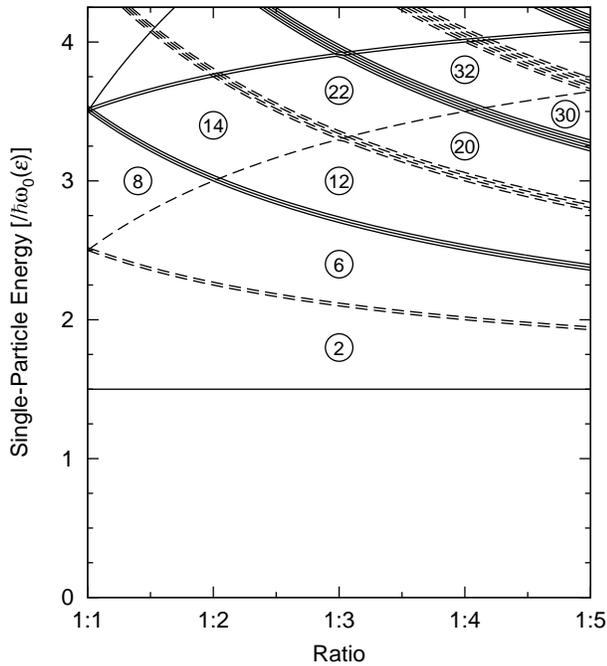} \caption{Nilsson
diagram versus the aspect ratio of the short (the $z$ direction) to long
(the radial direction) axes for an ellipsoidal nuclear surface (oblate
deformations).  The solid and dashed lines denote the single-particle
states with positive and negative parities, respectively.  The aspect
ratio 1:1 corresponds to the spherical shape.  The aspect ratio 1:5 is
close to that of a torus isomer obtained by the cranked HF calculation.
} \label{nil_dho}
\end{figure}

\subsection{Shell structure of torus nucleus}

\begin{figure*}[t]
\includegraphics[keepaspectratio,width=\linewidth]{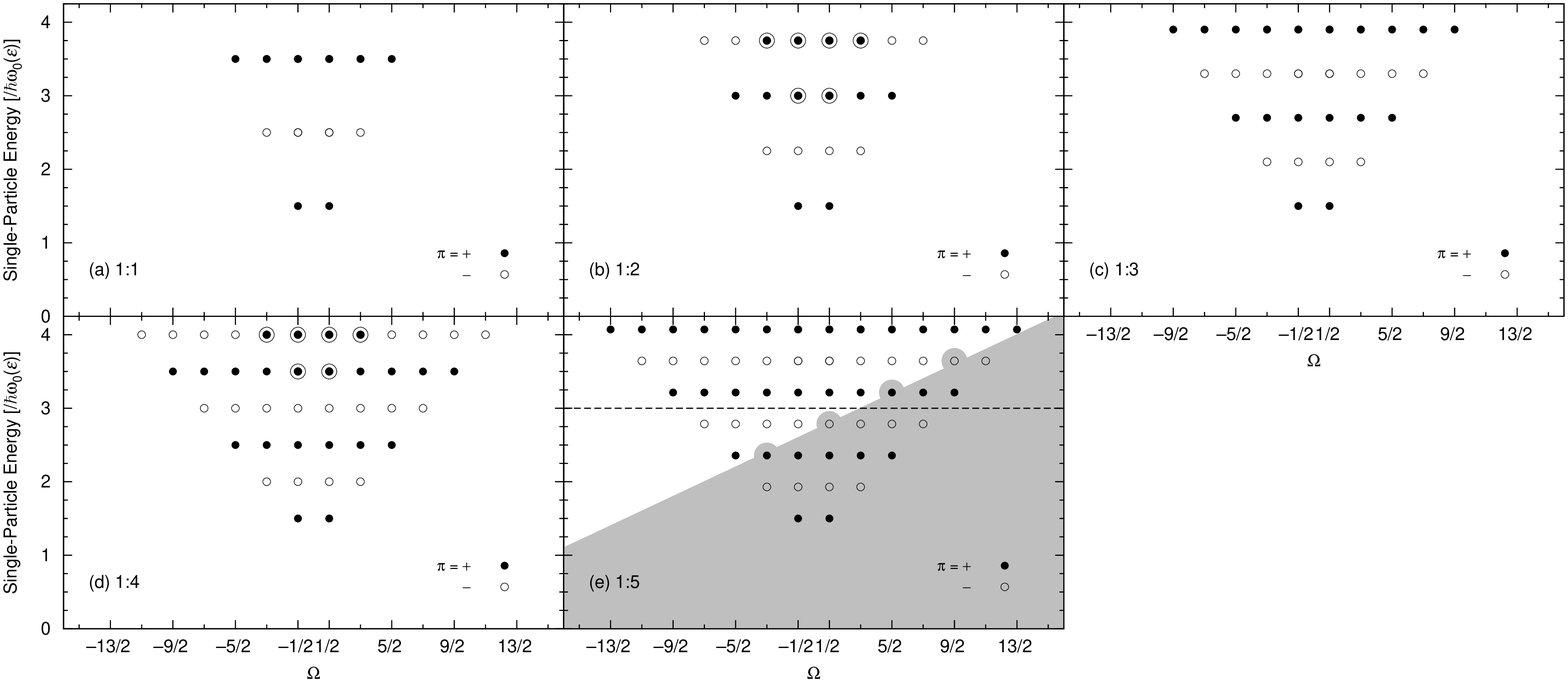}
\caption{Single-particle energies of the deformed harmonic-oscillator
versus $\Omega$ at various oblate deformations.  In panel (e), the
dashed line denotes the Fermi surface for $^{40}$Ca ($N=20$) at
$\omega=0$.  The gray area denotes the occupied states in $^{40}$Ca at
$\omega=1.6$ MeV/$\hbar$ for rotation about the symmetry axis. }
\label{spc_dho}
\end{figure*}

Using the RDHO model, we next investigate shell structures of a torus
isomer and examine how single-particle configurations change from
spherical to torus shapes.  Figure \ref{nilrdho} shows a Nilsson diagram
versus the parameter $\eta=R_0/d$ for $^{40}$Ca.  At $\eta=0$, the
nuclear shape is spherical.  At $\eta=4$, a torus shape is well
developed, which is a size similar to that obtained by the cranked HF
calculations.  Note that we take into account volume conservation inside
an equi-potential surface of a torus isomer (see Ref.~\cite{Wong73} and
Appendix for the volume conservation in $0\le \eta \le 1$).  To
eliminate the volume effect, we plot the single-particle energies in
unit of $\hbar\omega_0(\eta)$.  In this figure, we slightly shift the
single-particle energies with higher $\Omega$ in order to illustrate the
degeneracy of the states.

In Fig.~\ref{nilrdho}, we see the spherical major shell with
$E=\hbar\omega _0(N_{\rm sh}+3/2)$ at $\eta=0$, where $N_{\rm sh}$
denotes the total number of oscillator quanta.  With increasing $\eta$,
the single-particle energies with $\Omega=1/2$ approach the asymptotic
value given by $E=\hbar\omega_0(N'_{\rm sh}+1)$, where $N'_{\rm
sh}=n_r+n_z$, $n_z$ and $n_r$ denote the quantum number for oscillations
in the $z$ and the radial directions, respectively. The energies of
other single-particle states with larger $\Omega$ in the same $N_{\rm
sh}$ shell steeply decrease as a function of $\eta$.  At $\eta=4$, the
10th and 11th (from the bottom) single-particle states with
$\Omega^p=11/2^-$ and $13/2^-$ ($\Lambda=6$) become lower than the 14th
level with $N'_{\rm sh}=1$ (the 1$\hbar\omega_0$ state).  These two
single-particle states originate from those with a spherical
harmonic-oscillator quantum number of $N_{\rm sh}=5$ (the
$5\hbar\omega_0$ state).

It is easy to understand these behaviors.  As Wong showed in
Ref.~\cite{Wong73}, the single-particle energies for large $R_0$ are
approximately given by $E\sim \hbar\omega_0(N'_{\rm
sh}+1)+\hbar^2\Lambda^2/2mR_0^2$. Thus, the single-particle energies
belonging to the same $N'_{\rm sh}$ shell are proportional to
$\Lambda^2$ at lager $R_0$.

Figure \ref{spc_rdho} shows the single-particle energies in the RDHO
model versus $\Omega$ from $\eta=0$ to 4.  At $\eta=0$
[Fig.~\ref{spc_rdho} (a)], the familiar shell structure of the spherical
harmonic-oscillator is seen.  With increasing $\eta$
[Figs.~\ref{spc_rdho} (b)-(d)], single-particle energies with high
$\Omega$ rapidly decrease.  Then, the single-particle energies start to
form parabolic structures.  At $\eta=4$ [Fig.~\ref{spc_rdho} (e)], two
important properties emerge: (i) the curvature of the parabolic
structure becomes large, and (ii) the single-particle energies within
the same $N_{\rm sh}'$ shell are proportional to $\Lambda^2$.  These two
properties play an essential role in stabilizing the torus isomers when
single particles are aligned in the direction of the symmetry axis.

It is surprising that the single-particle shell structure of the RDHO
model at $\eta=4$ is very similar to that of Fig.~\ref{figsp3} obtained
by the cranked HF calculation.  The RDHO model is therefore a good
approximation for describing the microscopic structures of the torus
isomers.
 
\subsection{Emergence of the torus shape beyond the limit of oblate deformation} 
\begin{figure}[htbp]
\includegraphics[keepaspectratio,width=6cm]{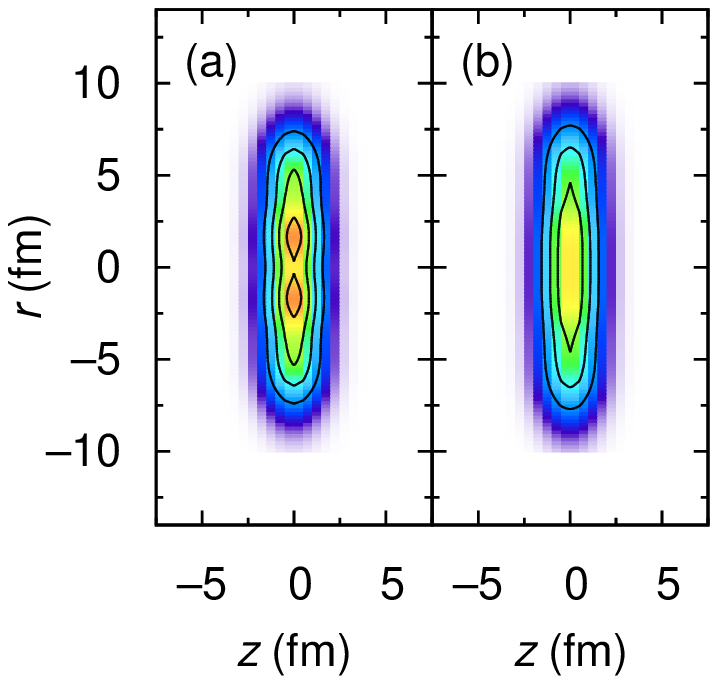} \caption{ (a)
 Density distributions of neutrons in $^{40}$Ca calculated for the
 deformed harmonic-oscillator model at an oblate deformation of the
 aspect ratio 1:5 with $\omega=0$.  The contours correspond to multiple
 steps of 0.05 fm$^{-2}$.  The densities for the $\phi$ direction in the
 cylindrical coordinate are integrated.  (b) The same as (a) but with
 $\omega=$1.6 MeV/$\hbar$.  The colors are normalized by the largest
 density of panel (a). } \label{den_dho}
\end{figure}
\begin{figure}[htbp]
\includegraphics[keepaspectratio,width=\linewidth]{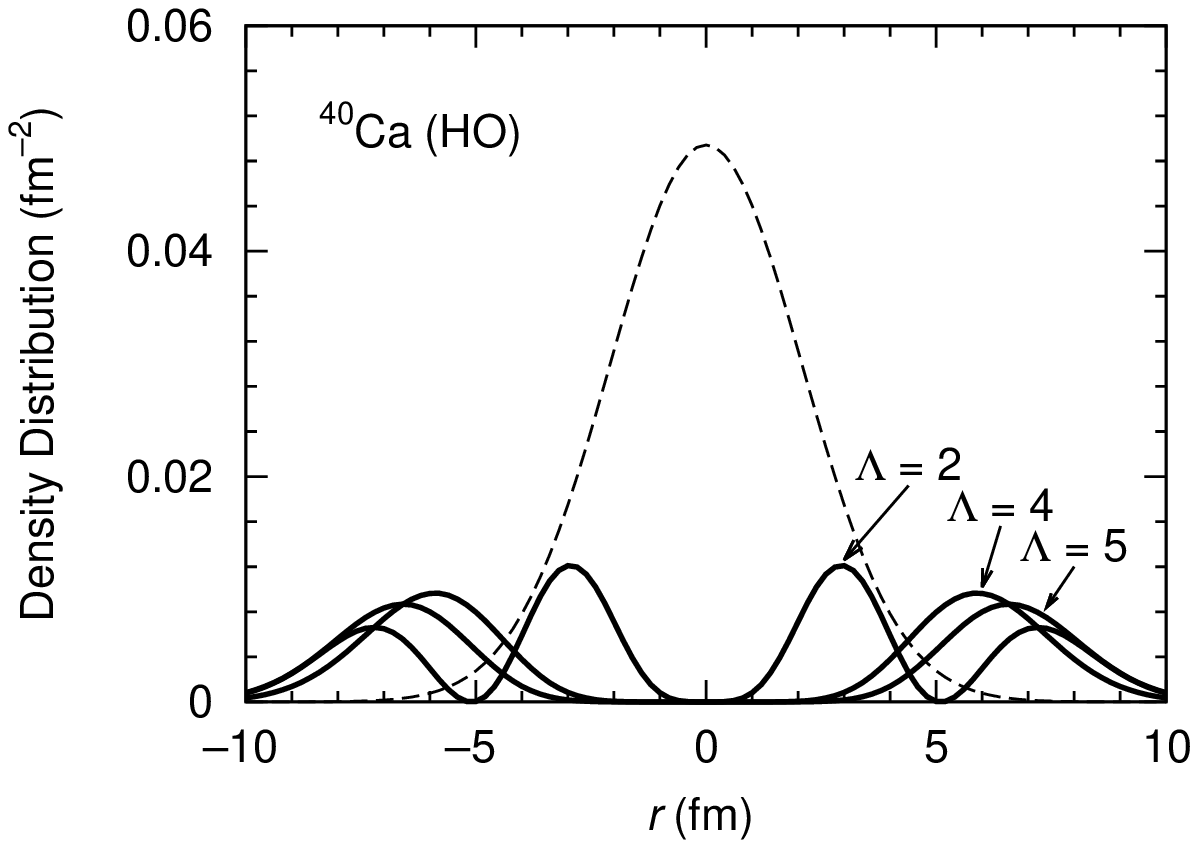}
\caption{Density distributions of the single-particle states of special
interest for $^{40}$Ca on the $z=0$ plane in an oblate deformation of
the aspect ratio 1:5 calculated with the deformed harmonic-oscillator
model. The densities in the $\phi$ direction in the cylindrical
coordinate are integrated. The solid lines show the density
distributions of the aligned single-particle states with $\Lambda=2$, 4, and
5 that are occupied at $\omega=1.6$ MeV/$\hbar$. 
The dashed line shows the density distribution of the lowest 
$\Lambda=0$ state.}  \label{wf_dho}
\end{figure}

Lastly, let us discuss the reason why the torus nucleus emerges beyond
the limit of large oblate deformation.  In the spherical
harmonic-oscillator potential, the radial wave function of the lowest
single-particle state is given by a Gaussian function peaked at the
center (the $0s$ state): accordingly, the central part of the total
density distribution is quite stable.  Then, a question arises why such
a stable and robust wave function vanishes and how the torus shape
emerges.

To investigate why the $0s$ state disappears, we calculate the
single-particle energies for the deformed harmonic-oscillator potential
as a function of oblate deformation.  Figure \ref{nil_dho} shows the
obtained Nilsson diagram versus the aspect ratio of the short (the $z$
direction) to the long (the radial direction) axes for an ellipsoidal
nuclear surface (oblate deformation).  The aspect ratio 1:1 corresponds
to the spherical shape. The aspect ratio 1:5 corresponds to an oblate
shape with the same aspect ratio as that of the torus isomer of
$^{40}$Ca obtained by the cranked HF calculation.  The single-particle
energies are plotted in unit of $\hbar\omega_0(\epsilon)$, where
$\omega_0(\epsilon)$ denotes the frequency of the harmonic-oscillator
potential depending on the Nilsson perturbed-spheroid parameter
$\epsilon$ to describe ellipsoidal nuclear shapes.  In
$\omega_0(\epsilon)$, the volume conservation inside an equi-potential
surface is taken into account \cite{Nilsson}.  In the figure, we see
that some single-particle energies associated with high $N_{\rm sh}$
spherical major shells rapidly decrease with increasing oblate
deformation.  At the aspect ratio 1:5, the last occupied state for
$^{40}$Ca ($N=20$) originates from that with a spherical
harmonic-oscillator quantum number of $N_{\rm sh}=3$ (the
$3\hbar\omega_0$ state).

In Figs.~\ref{spc_dho}, the single-particle energies are plotted versus
$\Omega$ at each aspect ratio.  We see that the shell gaps of the
single-particle energies decrease with increasing oblate deformation.
However, the basic pattern of deformed shell structure does not change,
in contrast to that of the RDHO model shown in Fig.~\ref{spc_rdho}.  In
Fig.~\ref{spc_dho} (e), the dashed line denotes the Fermi level for
$N=20$ at $\omega=0$. The neutron density distribution, $\rho(r,z)$, for
the occupied configuration is shown in Fig.~\ref{den_dho} (a). The
densities in the $\varphi$ direction are integrated. Two prominent peaks
are seen in the density distribution.  We next consider an aligned
single-particle configuration at $\omega=1.6$ MeV/$\hbar$. This $\omega$
corresponds to a value for the torus isomer of $^{40}$Ca.  The occupied
states at this $\omega$ are shown by the gray area in Fig.~\ref{spc_dho}
(e).  By the alignments, totally five single-particle states with
$11/2^-$[505] and $9/2^-$[505] ($\Lambda=5$), $9/2^+$[404] and
$7/2^+$[404] ($\Lambda=4$), and $5/2^+$[402] ($\Lambda=2$) are occupied
(the asymptotic Nilsson label $\Omega^\pi[Nn_z\Lambda]$ is used here).
On the other hand, the single particle states with $-7/2^-[303]$ and
$-5/2^-[303]$ ($\Lambda=3$), $-5/2^+[202]$ ($\Lambda=2$), and
$-3/2^-[301]$ and $-1/2^-[301]$ ($\Lambda=1$) become unoccupied.
Summing up the aligned single-particle angular momenta, we obtain the
neutron contribution to the $z$ component of the total angular momentum
$J_z=31$ $\hbar$.  Taking into account the proton contribution as well,
we finally obtain the total angular momentum $J_z=62$ $\hbar$ for this
oblate configuration, which is close to that of the torus isomer for
$^{40}$Ca obtained by the cranked HF calculation.  The neutron density
distribution at $\omega=1.6$ MeV/$\hbar$ is shown in Fig.~\ref{den_dho}
(b).  The two peaks seen in Fig.~\ref{den_dho} (a) vanish and densities
in the central region become flat and stretch to radial direction, as
the single-particle states with high $\Omega$ are occupied.

Figure \ref{wf_dho} shows the density distributions of individual
single-particle states of special interest at aspect ratio 1:5.  The
densities for the $\varphi$ direction in the cylindrical coordinate are
integrated. The dashed line shows the density distribution of the lowest
$\Lambda=0$ state. On the other hand, the solid lines depict those of
the aligned $\Lambda=2$, 4, and 5 states mentioned above that are
occupied at $\omega=1.6$ MeV/$\hbar$.  The single-particle density
distributions of these aligned states peak around $r=6$ fm.  Apparently,
the overlap between the aligned nucleons and the nucleons in the lowest
$\Lambda=0$ state is very small.  Namely, the lowest $\Lambda=0$ state
largely containing the spherical $0s$ component is rather isolated from
the others.  To gain the attractive interactions between nucleons, the
total system tends to maximize the overlaps between the density
distributions of individual single particles.  Thus, it would be
energetically favorable to concentrate the densities of individual
nucleons around $r=6$ fm.  In this way, the nucleus with extremely large
oblate deformation may start to generate the torus shape. This seems to
be the basic reason why a large 'hole' is created in the central region
of the nucleus by eliminating the spherical $0s$ wave function.

\section{Summary}
We have systematically investigated the existence of high-spin torus
isomers for a series of $N=Z$ even-even nuclei from $^{28}$Si to
$^{56}$Ni using the cranked HF method.  We found the stable torus
isomers from $^{36}$Ar to $^{52}$Fe for all the Skyrme interactions used
in this study.  In the obtained torus isomers, the $z$ components of the
total angular momentum are $J_z=36$ $\hbar$ for $^{36}$Ar, 60 $\hbar$
for $^{40}$Ca, 44 $\hbar$ for $^{44}$Ti, 72 $\hbar$ for $^{48}$Cr, and
52 $\hbar$ for $^{52}$Fe.  We fitted the density distribution of each of
the obtained torus isomers with the Gaussian function.  We also analyzed
the microscopic structure of the obtained torus isomers by plotting the
single-particle energies versus $\Omega$ and using the concept of
sloping Fermi surface. We determined the regions of $\omega$ for which
the obtained torus isomers can stably exist in each Skyrme interaction.
The dependence of the obtained results on the Skyrme interactions
employed is found to be weak.

We have also performed TDHF calculations to explore the properties of
the precession motion rotating with angular momentum $I=K+1$, which is
built on each of the obtained torus isomer with $I=K$.  For all the
obtained torus isomers, the precession motion emerges and the symmetry
axis rotates about the precession axis for at least two periods.  It was
found that the precession motion of the 60 $\hbar$ isomer in $^{40}$Ca
is especially robust and stably rotates for many periods, We obtained
similar results for all the Skyrme interactions used, We also
estimated the moment of inertia for the rotation about a perpendicular
axis from the calculated rotational periods of the precession motion.
The obtained moments of inertia are close to the rigid-body values for
all the obtained torus isomers.

We have discussed the radial density distribution of each
single-particle wave function in the high-spin torus isomer of
$^{40}$Ca. We showed that the density distributions are well
approximated by those of the RDHO model.  We then discussed how the
shell structure develops from spherical to torus shapes.  There are two
important mechanisms for stabilizing torus isomers: (i) the development
of the major shells consisting of single-particle states whose energies
are given by $E=(N_{\rm sh}'+1)+\hbar^2\Lambda^2/2mR_0^2$, where $N_{\rm
sh}'=n_r+n_z$ and $\Lambda$ is the $z$ component of orbital angular
momentum. (ii) a large value of $R_0$ that reduces the energies of high
$\Omega$ single-particle states.  We finally discussed why the $0s$
components of all the single-particle wave functions vanish and generate
a torus shape. We showed that in an aligned single-particle
configuration with extremely large oblate deformation, the overlaps
between the density distributions of the lowest $\Lambda=0$
single-particle state and the aligned high-$\Omega$ single-particle
states become very small due to the strong centrifugal force.  To gain
the attractive interaction energy as much as possible, nucleons tend to
maximize the overlaps of their wave functions.  An optimal configuration
beyond the limit of large oblate deformation is the one creating the
localization of single-particle density distributions around a torus
ring.  This seems to be a basic mechanism of the emergence of high-spin
torus isomers.

\begin{acknowledgments}
 TI was supported in part by MEXT SPIRE and JICFuS.
 This work was undertaken as part by the Yukawa International Project
 for Quark-Hadron Sciences (YIPQS).  J.A. M. was supported by BMBF under
 contract number 06FY9086 and 05P12RFFTG, respectively.
\end{acknowledgments}

\appendix*
\section{PARAMETERS AND VOLUME CONSERVATION IN THE RDHO MODEL} 
In the RDHO model, we take the oscillator frequency, $\omega_0$, to
conserve the inner volume of an equi-potential energy surface.  It is
given by
\begin{equation}
\left(\frac{\omega_0}{\stackrel{\circ}{\omega}_0}\right)^3 =
   \begin{cases}
    \begin{split}
     (1&+\frac{\eta^2}{2})\sqrt{1-\eta^2} \\
     &+\frac{3}{4}\pi\eta(1+\frac{2}{\pi}
     \arctan\frac{\eta}{\sqrt{1-\eta^2}})
    \end{split}
     & (0 \le \eta < 1) \\
    \frac{3}{2}\pi\eta & (\eta\ge1),
   \end{cases}
\end{equation}
where $\eta=R_0/d$ and $\stackrel{\circ}{\omega}_0$ denotes the
oscillator frequency in the spherical limit. Here, we take
$\hbar\!\!\stackrel{\circ}{\omega}_0=41 A^{-1/3} \rho_{\rm
torus}/\rho_{\rm gr}$ MeV, where $A$ is the number of nucleons, and
$\rho_{\rm torus}$ and $\rho_{\rm gr}$ denote the average densities of a
torus isomer and the ground state, respectively~\cite{Stas14}.  In the
calculations, we use $\rho_{\rm torus}=(2/3)\rho_{\rm gr}$.

\end{document}